\documentclass[aps,pre,onecolumn,floatfix]{revtex4}
\usepackage{graphicx, bm, bbm,amsmath, amssymb, epsfig,color,slashbox}

\pdfoutput=1

\newcommand{\nn}{\noindent}

\newcommand{\bq}{\begin{align}}
\newcommand{\eq}{\end{align}}

\begin{document}
\title{Efficient bending and lifting patterns in snake locomotion}

\author{Silas Alben$^*$}
%\thanks{}
\affiliation{Department of Mathematics, University of Michigan,
Ann Arbor, MI 48109, USA}
\email{alben@umich.edu}

\date{\today}

\begin{abstract}
We optimize three-dimensional snake kinematics for locomotor efficiency. We assume a general space-curve representation of the snake backbone with small-to-moderate lifting off the ground and negligible body inertia. The cost of locomotion includes work against friction and internal viscous dissipation. When restricted to planar kinematics, our population-based optimization method finds the same types of optima as a previous Newton-based method. A few types of optimal motions prevail. We find an s-shaped body with alternating lifting of the middle and ends for small-to-moderate transverse friction. For large transverse friction, curling and sliding motions are typical with small viscous dissipation, replaced by large-amplitude bending with large viscous dissipation. With small viscous dissipation we find local optima that resemble sidewinding motions across friction coefficient space. They are always suboptimal to alternating lifting motions, with average input power 10--100\% higher.
\end{abstract}


\maketitle

\section{Introduction}

Snakes have a relatively simple body geometry but can perform a wide range of locomotor behaviors. They are useful for understanding the mechanics of locomotion in terrestrial and even aquatic \cite{shine2003aquatic} and aerial \cite{socha2002kinematics} environments. 
Snakes are also an important source of ideas for bioinspired robots \cite{hirosebiologically,transeth2009survey,hopkins2009survey,liljeback2012snake,HaCh2010b,astley2015modulation}, and their limblessness has advantages for control, adaptability, and navigation in complex and cluttered environments \cite{fu2020robotic,astley2020side,fu2022snakes}. Although a wide range of motions are possible, four major modes of snake locomotion---serpentine, concertina, sidewinding, and rectilinear---have been described and studied most often \cite{gray1946mechanism,gans1970snakes,jayne1986kinematics,lillywhite2014snakes}, though the true diversity of motions is greater \cite{gans1984slide,transeth2009survey,jayne2020defines}.

Many common biological snake motions such as serpentine locomotion have been modeled successfully by assuming planar motions with a simple (Coulomb) frictional model \cite{ma2001analysis,sato2002serpentine,chernousko2005modelling,GuMa2008a,HuNiScSh2009a,HuSh2012a,aguilar2016review,yona2019wheeled,rieser2021functional}.
Serpentine and concertina-like motions were found to be optimally efficient among general time-periodic kinematics of
three-link \cite{alben2019efficient} and smooth bodies \cite{AlbenSnake2013,wang2014optimizing}. For certain body geometries and frictional anisotropies, other motions, beyond those observed biologically, were found to be optimal \cite{JiAl2013,alben2020intermittent,alben2021efficient}.

These models assume a resistive force law, local in the velocity, which is somewhat simpler than many fluid locomotion models \cite{childress1981mechanics,sparenberg1994hydrodynamic,alben2008optimal,alben2009swimming}. Nonetheless, 
even the simplest planar models with Coulomb friction are too complicated to be solved theoretically with large-amplitude motions, so the physics of many motions are not well understood, and computational models and explorations of the types of efficient dynamics over parameter space play an important role. 

Of the four major modes of locomotion, nonplanar kinematics feature most strongly in sidewinding \cite{gray1946mechanism,jayne1986kinematics,marvi2014sidewinding}. \cite{marvi2014sidewinding,astley2015modulation} demonstrated that sidewinding can be represented as a pair of orthogonal body waves (vertical and horizontal) that can be independently modulated to achieve high maneuverability and ascend inclines. The other major modes are not perfectly planar in real snakes or robots but can be approximated well by bodies that remain planar. The serpentine mode often features lifting at the curvature peaks, termed ``sinus-lifting" \cite{hirosebiologically,transeth2009survey}, and modeled by \cite{HuNiScSh2009a,HuSh2012a} using planar shapes together with a weight distribution or lifting function. 
\cite{zhang2021friction} used planar shapes together with the weight distribution function approach of \cite{HuNiScSh2009a,HuSh2012a} to model sidewinding and other three-dimensional (3D) motions. They used a sinusoidal weight distribution function and a planar body curvature with a phase shift between them, and examined the dynamics across the parameter space of the relative phase shift, the weight distribution function amplitude, and the ratio of lateral to lifting wave numbers. They found a wide range of turning, slithering, and sidewinding motions. In a test case, the planar model with a weight distribution was consistent with a direct 3D simulation using a Cosserat rod model. 
\cite{rieser2021functional} used sinusoidal curvature and weight distribution functions to model sidewinding locomotion. They showed that sidewinding motion travels farther per period with isotropic friction than with transverse-friction-dominated anisotropy, which is useful for lateral undulation.  

Here, instead of a planar curve with a weight distribution function, we represent the snake body as a time-dependent 3D space curve. Given the space curve shape, the regions of contact and lifting arise naturally from solving the dynamical equations. We prefer this direct approach because it removes the question of which 3D shape, if any, would produce a prescribed weight distribution function. The weight distribution depends on the equilibrium position of the space curve under gravity and may be sensitive to slight changes in the curve's shape.

Three-dimensionality opens up new possibilities for interesting mechanisms of locomotion. Some 3D motions may have higher efficiency than similar two-dimensional (2D) motions, e.g. sinus-lifting versus lateral undulation, and 3D motions may be necessary for traversing uneven ground and obstacles. Some 3D studies have focused on modifications of sidewinding motions: \cite{astley2020side} showed how sidewinding waveforms can be modulated in biological and robotic snakes to move around peg obstacles and \cite{chong2022moving} developed a geometric mechanics method to find optimal contact patterns for sidewinding robotic snakes for maximum speed. Other types of 3D motions have also been studied: \cite{fu2022snakes} studied experimentally how biological snakes used a combination of lateral and vertical bending to traverse uneven terrain made up of blocks with vertical and horizontal planar surfaces; \cite{fu2020robotic} showed that body compliance can help snakes traverse obstacles stably. 

As in our previous planar locomotion studies \cite{JiAl2013,AlbenSnake2013,wang2014optimizing,alben2019efficient,alben2020intermittent,alben2021efficient}, 
we study nonplanar motions using an optimization framework. It is not feasible to describe the full range of locomotor behaviors across the space of geometrical and physical parameters. Focusing on those that are optimally efficient is more manageable, even if they do not fully describe the possibilities. The optimal solutions indicate the tradeoffs between the objectives and constraints in the problem, and how they depend on the parameters \cite{alexander1996optima}. The optimal solutions can provide effective strategies and suggest general mechanisms for robotic locomotion \cite{hopkins2009survey,chong2022moving}. The relationship between the optima and biological organisms is less clear, as the model omits many physiological aspects, and the importance of mechanical efficiency to reproductive success varies widely among organisms \cite{langerhans2010ecology}. Comparing the optimal solutions to biological locomotion indicates the importance of mechanical efficiency relative to other factors in determining the choice of locomotor mode \cite{alexander1996optima}.

In \cite{AlbenSnake2013,alben2020intermittent} we used quasi-Newton (BFGS) optimization methods; in \cite{AlbenSnake2013,wang2014optimizing} we used theoretical methods applicable to small body deflections; in 
\cite{JiAl2013} we used a commercial solver; and in \cite{alben2021efficient} we used a stochastic population-based optimization method. Here we also use a stochastic population-based method, because it is relatively simple and turns out to be effective and robust for the problem at hand. Convergence is generically slower than for Newton-based optimization methods, but those methods may have difficulties because the normal contact and tangential Coulomb friction forces where the body meets the ground are singular (or nearly singular in the regularized model we use), which also causes singularities in derivatives of the objective function.

Section \ref{sec:model} describes our model of a locomoting snake---a curve in 3D space with curvature and torsion that vary with arc length and time. The curve moves according to Newton's laws, with forces due to friction and normal contact with a planar ground. Unlike our previous optimization studies of 2D motions, here we include two terms in the cost function---the rate of power dissipation by the frictional force at the ground plus a new term, the internal viscous power dissipation for a linearly viscoelastic rod (the Kelvin-Voigt model). We solve this nonlinear model using an implicit time-stepping approach. Section \ref{sec:Stochastic} describes our optimization method. We compute the motions of a population of locomoting bodies, select a top-performing subset, and use random perturbations of its members to form the population at the next generation. After many generations, the population converges to the vicinity of a local optimum.

Section \ref{sec:results} describes the computed optima and how they vary with key parameters such as the numbers of modes describing the curvature and torsion (section \ref{sec:modes}). Section \ref{sec:planar} shows how the algorithm behaves in the special case of planar motions, both with and without internal viscosity. In the latter case good agreement with \cite{AlbenSnake2013} is found. Section \ref{sec:lifting} then shows how the optimal motions change as the
limit on the amount of lifting/nonplanarity increases from zero. With a small to moderate amount of allowed lifting, the
effect of frictional anisotropy and damping on the optimal motions is studied in section \ref{sec:anisotropies}. A special class of optima that resemble sidewinding motions are shown in section \ref{sec:sidewinding}, and special types of analytical optima are presented in section \ref{sec:theoretical}. Section \ref{sec:conclusions} gives the conclusions.

\section{Model \label{sec:model}}

\begin{figure} [h]
           \begin{center}
           \begin{tabular}{c}
 \hspace{-0.2in}
               \includegraphics[width=5.5in]{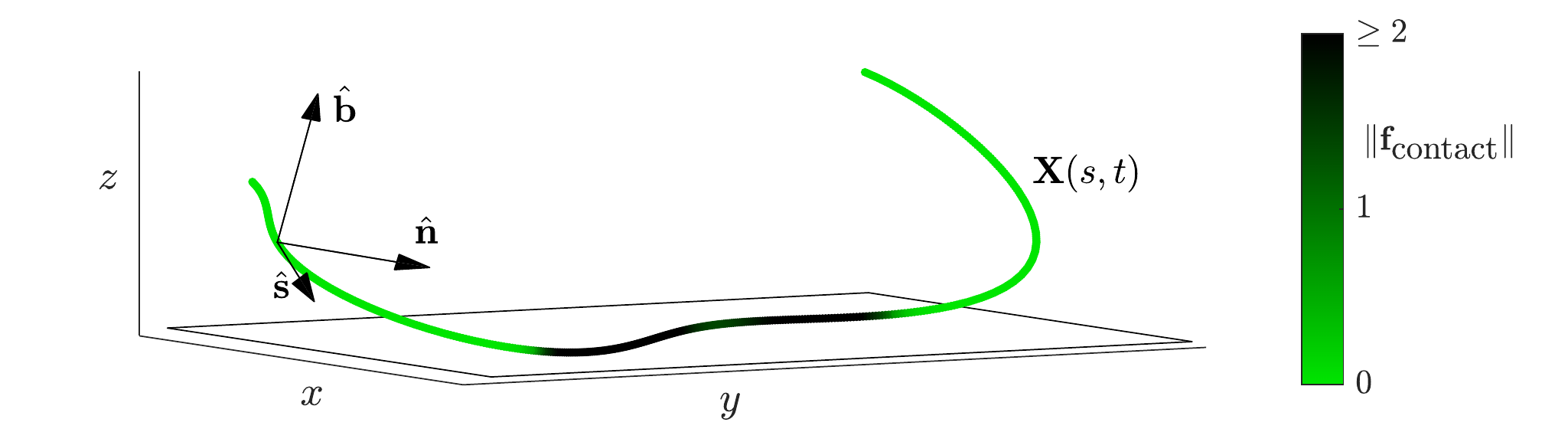} \\
           \end{tabular}
          \caption{\footnotesize Schematic diagram of the 3D space curve $\mathbf{X}(s,t)$ representing the snake backbone, together with the tangent, normal, and binormal vectors $\{\hat{\mathbf{s}}, \hat{\mathbf{n}}, \hat{\mathbf{b}}\}$ at one location. The curve rests on the plane $z = 0$, the ground, indicated by a rectangle. The color scale shows the magnitude of the ground contact force acting at each location.  
 \label{fig:Schematic3DSnake}}
           \end{center}
         \vspace{-.10in}
        \end{figure}

In previous work, we assumed that the snake body was a curve in the 2D plane, with a prescribed shape given by its curvature $\kappa(s,t)$, where $s$ is arc length and $t$ is time. Now, we assume the body is a curve in 3D space (see figure \ref{fig:Schematic3DSnake}), so the shape is given by its curvature $\kappa(s,t)$ and torsion $\tau(s,t)$ \cite{GuggenDG}. We assume that $\kappa(s,t)$ and $\tau(s,t)$ are periodic in time with period $T$. Given $\kappa$ and $\tau$, we can integrate the Frenet-Serret formulas
\begin{align}
\frac{d\hat{\mathbf{s}}}{ds} & = \kappa \hat{\mathbf{n}} \quad ; \quad
\frac{d\hat{\mathbf{n}}}{ds} = -\kappa \hat{\mathbf{s}} + \tau \hat{\mathbf{b}} \quad ; \quad
\frac{d\hat{\mathbf{b}}}{ds} = -\tau \hat{\mathbf{n}}
\end{align}
\nn to obtain the body tangent, normal, and binormal vectors $\{\hat{\mathbf{s}}(s,t), \hat{\mathbf{n}}(s,t), \hat{\mathbf{b}}(s,t)\}$, if we know these vectors at one $s$ value, say $s = 0$ (the tail). The values of these vectors at the tail can be written as the columns of a 3D rotation matrix $\left[\hat{\mathbf{s}}(0,t) \;\;\hat{\mathbf{n}}(0,t)\;\;\hat{\mathbf{b}}(0,t)\right] = \mathbf{R}_z\left(\alpha(t)\right) \mathbf{R}_y\left(\beta(t)\right)\mathbf{R}_x\left(\gamma(t)\right)$, where $\mathbf{R}_z$, $\mathbf{R}_y$, and $\mathbf{R}_x$ are the matrices for counterclockwise rotations about the $z$, $y$, and $x$ axes by (Euler) angles $\alpha(t)$, $\beta(t)$, and $\gamma(t)$, respectively \cite{GuggenDG}.

%\begin{equation}
%\left[\hat{\mathbf{s}}(0,t) \;\;\hat{\mathbf{n}}(0,t)\;\;\hat{\mathbf{b}}(0,t)\right] =\begin{bmatrix}\cos \alpha \cos \beta \;\;&\;\;\cos \alpha \sin \beta \sin \gamma -\sin \alpha \cos \gamma \;\;&\;\;\cos \alpha \sin \beta \cos \gamma +\sin \alpha \sin \gamma \\ 
%\sin \alpha \cos \beta \;\;&\;\; \sin \alpha \sin \beta \sin \gamma +\cos \alpha \cos \gamma \;\;&\;\;\sin \alpha \sin \beta \cos \gamma -\cos \alpha \sin \gamma \\
%-\sin \beta \;\;&\;\;\cos \beta \sin \gamma \;\;&\;\;\cos \beta \cos \gamma \end{bmatrix}
%\end{equation}
%\nn given by three parameters, 
%$\{\alpha(t), \beta(t), \gamma(t)\}$, also known as Euler angles \cite{GuggenDG}. 
After integrating the Frenet-Serret formulas, we have 
$\hat{\mathbf{s}}(s,t) = \partial_s\mathbf{X}(s,t)$, where $\mathbf{X}(s,t) = [x(s,t),y(s,t),z(s,t)]^T$ is the position of the body. Given the position of the body at the tail, $\mathbf{X}_0(t) \equiv \mathbf{X}(0,t)$, we integrate $\hat{\mathbf{s}}$ to obtain the position all along the body,
\begin{align}
\mathbf{X}(s,t) = \mathbf{X}_0(t) + \int_0^s \hat{\mathbf{s}}(s',t) ds'.
\end{align}
\nn To summarize, given the shape of the body ($\kappa$ and $\tau$), and the six unknowns $\{\mathbf{X}_0(t), \alpha(t), \beta(t), \gamma(t)\}$ specifying its position and orientation at the tail, we can integrate to obtain $\mathbf{X}(s,t)$. At each $t$, we solve for the six unknowns by enforcing six equations, the linear and angular momentum balances for the snake body as a whole:
\begin{align}
\rho \int_0^L \partial_{tt} \mathbf{X} ds &= \int_0^L \mathbf{f}_{\mbox{\scriptsize ext}} ds \quad ; \quad
\partial_t \left( \rho \int_0^L \mathbf{X} \times \partial_{t} \mathbf{X} ds \right) = \rho \int_0^L \mathbf{X} \times \partial_{tt} \mathbf{X} ds = \int_0^L \mathbf{X} \times \mathbf{f}_{\mbox{\scriptsize ext}} ds. \label{LinAngMom}
\end{align}
\nn Here $L$ is the body length, $\rho$ is the mass per unit length of the body and $\mathbf{f}_{ext}$ is the external force on the body, due to normal contact with the ground, gravity, and friction:
\begin{align}
\mathbf{f}_{\mbox{\scriptsize ext}} &= \mathbf{f}_{\mbox{\scriptsize contact}} + \mathbf{f}_{\mbox{\scriptsize gravity}} + \mathbf{f}_{\mbox{\scriptsize friction}} \quad ; \quad \mathbf{f}_{\mbox{\scriptsize contact}} = \rho g \mathbf{e}_z \left(H(z)e^{- z/\delta_w} + (1-H(z)) (1 - z/\delta_w) \right) \label{fcontact}\\
\mathbf{f}_{\mbox{\scriptsize gravity}} &= -\rho g \mathbf{e}_z \quad ; \quad
\mathbf{f}_{\mbox{\scriptsize friction}} = -\| \mathbf{f}_{\mbox{\scriptsize contact}}\| \frac{\mu_s \left(\partial_t \mathbf{X}_{2D}\cdot \hat{\mathbf{s}}_{2D}\right) \hat{\mathbf{s}}_{2D} + \mu_n \left(\partial_t \mathbf{X}_{2D}\cdot \hat{\mathbf{s}}_{2D}^\perp\right) \hat{\mathbf{s}}_{2D}^\perp}{\sqrt{\|\partial_t \mathbf{X}_{2D}\|^2+\delta_s^2}}. \label{fgravityfriction}
\end{align}
\nn Here $g$ is gravitational acceleration and $H$ is the Heaviside function. Instead of a hard or rigid contact force that rises in magnitude from 0 to $+\infty$ when the body penetrates the ground at $z = 0$, the contact force magnitude is very small when $z \gg \delta_w$ and very large when $z \ll -\delta_w$. We set $\delta_w = 10^{-3}L$, approximating the hard contact limit. Making $\delta_w$ much smaller than $10^{-3}L$ does not noticeably alter the body dynamics in our computations but can significantly increase the number of iterations required by the iterative solver of the nonlinear ODE system (\ref{LinAngMom}).

The frictional force model is an extension of the 2D version used in several previous works \cite{ma2001analysis,sato2002serpentine,chernousko2005modelling,GuMa2008a,HuNiScSh2009a,alben2019efficient}. Sliding friction opposes the component of velocity tangent to the ground, written here as $\partial_t \mathbf{X}_{2D}$, the projection of the body velocity in the $x$-$y$ plane. 
The frictional force magnitude is proportional to the contact force magnitude and approximately independent of the velocity magnitude, following the Coulomb friction model.
We allow for anisotropic friction that corresponds to easier sliding in certain directions. 
For example, snake scales allow for smaller friction when the snake slides toward the head, in the $\hat{\mathbf{s}}$ direction, instead of toward the tail or perpendicular to the body axis \cite{sato2002serpentine,HuNiScSh2009a}. Stronger anisotropies can occur in robotic snakes due to wheels or active scales on the body surface \cite{hirosebiologically,MaHu2012a,aguilar2016review}. In (\ref{fgravityfriction}) the friction coefficients $\mu_s$ and $\mu_n$ are used for the components of velocity that are parallel to the ground and tangent to the backbone, or perpendicular to the backbone, respectively.
Here $\hat{\mathbf{s}}_{2D}$ is the projection of $\hat{\mathbf{s}}$ in the $x$-$y$ plane, so the frictional force acts tangentially to the ground, and its magnitude decreases as $\hat{\mathbf{s}}$ becomes more vertical at a contact. Typically $\hat{\mathbf{s}} \approx \hat{\mathbf{s}}_{2D}$ at a contact on the body's interior, because the body is smooth and almost tangent to the ground at a contact (otherwise it would penetrate the ground). At the ends of the body, $\hat{\mathbf{s}}$ may be somewhat more vertical, though in the computations we will limit the nonplanarity of the snake body, so $\hat{\mathbf{s}} \approx \hat{\mathbf{s}}_{2D}$ at the ends also. In (\ref{fgravityfriction}) $\hat{\mathbf{s}}^\perp_{2D}$ is $\hat{\mathbf{s}}_{2D}$ rotated 90 degrees counterclockwise about the $\mathbf{e}_z$-axis. Thus $\hat{\mathbf{s}}^\perp_{2D}$ points in the $x$-$y$ plane, in the direction transverse to the body tangent, and $\mu_n$ gives the coefficient of transverse friction. Note that $\hat{\mathbf{s}}^\perp_{2D} \neq \hat{\mathbf{n}}$, as $\hat{\mathbf{n}}$ points in the direction towards which the body curves, which could be vertical even when $\hat{\mathbf{s}}$ lies in the $x$-$y$ plane.

The parameter $\delta_s$ in (\ref{fgravityfriction}) is set to $10^{-3}L/T$, and smoothes a discontinuity in the frictional force that would occur when
$\|\partial_t \mathbf{X}\| = 0$, if $\delta_s$ were zero. Like $\delta_w$, $\delta_s$ is used to make the iterative solver more robust without
noticeably altering the dynamics. In \cite{alben2019efficient} we noted that there are certain motions for which $\delta_s \neq 0$ is required for a solution to exist, but these are somewhat uncommon.  The tangential friction coefficient $\mu_s$ takes the values $\mu_f$ or $\mu_b$ when the tangential body motion is in the forward or backward direction, respectively: $\mu_s =
\mu_f H(\partial_t \mathbf{X}_{2D}\cdot \hat{\mathbf{s}}_{2D}) +
\mu_b \left((1 - H(\partial_t \mathbf{X}_{2D}\cdot \hat{\mathbf{s}}_{2D})\right)$.
Without loss of generality, we may assume $\mu_b \geq \mu_f$. 
Our focus here is on efficient body kinematics, with efficiency defined similarly to previous locomotion studies \cite{childress1981mechanics,sparenberg1994hydrodynamic,schultz2002power,AlbenSnake2013,wang2018dynamics}: among the body kinematics that result in a time-averaged center-of-mass speed $V$, we find the one(s) that minimize the time-averaged power consumption. The power consumption is the rate of work done by the body against gravity, contact, and frictional forces, and we also include the rate of internal viscous dissipation, assuming a linear viscoelastic model that we now describe. 

So far, we have only described the centerline of the body, as a space curve with torsion $\tau$ and curvature $\kappa$. We now discuss elastic deformations of the entire body around the centerline, assuming one of the simplest models for its internal mechanics. We take the deformation to be that of a Kirchhoff rod, with elastic deformation due to bending and twisting about the centerline, but with negligible extension of the rod centerline and with cross-sections remaining planar and normal to the centerline \cite{o2017modeling}. 
The stress tensor is a linear combination of the strain and rate-of-strain tensors, following the Kelvin-Voigt model \cite{linn2013geometrically}:
\begin{align}
\sigma_{ij} = E\epsilon_{ij}+\eta \partial_t \epsilon_{ij} 
\end{align}
\nn with $E$ the Young's modulus and $\eta$ the shear viscosity. 

In general, the cross-sectional shape may vary along the rod, and at each centerline location $s$, it has a 2-by-2 area-moment-of-inertia tensor that depends on how material is distributed in the cross-section \cite{o2017modeling}. Let $\mathbf{d}_1$ and $\mathbf{d}_2$ be the principal axes (eigenvectors) of the tensor at location $s$, with corresponding principal area moments of inertia (eigenvalues) $I_1$ and $I_2$. Then the rod's elastic energy per unit length at $s$ is given by 
\begin{align}
u = \frac{1}{2}\left(EI_1\nu_1^2 + EI_2\nu_2^2 + GI_3\nu_3^2\right). \label{elastic}
\end{align}
\nn using the Cosserat rod model of \cite{linn2013geometrically} specialized to the Kirchhoff-rod case of zero centerline extension and zero shear. Here $E$ is the Young's modulus and $\nu_1$ and $\nu_2$ are the principal curvatures, also called material curvatures. They measure the components of the curvature vector $\kappa\hat{\mathbf{n}}$ in the $\mathbf{d}_1$ and $\mathbf{d}_2$ directions. $GI_3$ is the torsional or twisting rigidity, with $G$ the shear modulus and $I_3$ the polar area-moment-of-inertia. $\nu_3$ is the material twist (different from the geometric torsion $\tau$), which quantifies the rate of rotation of the rod cross-section with respect to change of arc length along the centerline. 

The rate of viscous dissipation per unit length is
\begin{align}
v = \frac{1}{2}\left(\eta_E I_1\left(\partial_t \nu_1\right)^2 + \eta_E I_2\left(\partial_t \nu_2\right)^2 + \eta I_3 \left(\partial_t \nu_3\right)^2\right) \quad ; \quad  \eta_E = \zeta\left(1-2\nu\right)^2 + \frac{4}{3} \eta\left(1+\nu\right)^2. \label{dissipation}
\end{align}
\nn with $\eta_E$, the extensional viscosity, a function of the bulk viscosity $\zeta$, the shear viscosity $\eta$, and the Poisson ratio $\nu$ \cite{linn2013geometrically}.
For incompressible materials, $\nu = 1/2$ and $\eta_E = 3\eta$. The relation between $\nu_1$, $\nu_2$, $\nu_3$ and $\kappa$ and $\tau$ is given by 
%(OReilly 2017 chapter) 
Bonnet's theorem and Meusnier's theorems \cite{o2017modeling,love1892treatise}:
\begin{align}
\nu_1 &= \kappa \sin{\phi} \quad ; \quad \nu_2 = \kappa \cos{\phi} \quad ; \quad 
\nu_3 = \tau + \partial_s\phi \label{meusnier}
\end{align}
\nn where $\phi$, called the angle of twist, is the angle between $\hat{\mathbf{n}}$ and $\mathbf{d}_2$. In order to minimize the rate of viscous dissipation, 
we can make the $\partial_t\nu_3$-term in (\ref{dissipation}) zero
always, by making the material twist $\nu_3 \equiv 0$. For a given centerline shape with geometric torsion $\tau$, this is done by choosing the angle of twist so that $\partial_s \phi = -\tau$ in (\ref{meusnier}). Physiological constraints would prevent large twisting in a biological snake, but not in a robot. In any case, here we will confine ourselves to small-to-moderate $\tau$. Zero $\nu_3$ is then achieved with small-to-moderate $\partial_s \phi$, which is more feasible physiologically. For simplicity, we also assume $I_1 = I_2 = I$, as occurs for example with a rod of circular cross-section. Combining these assumptions, $v$ in (\ref{dissipation}) becomes
$\frac{1}{2} \eta_E I\left(\partial_t \kappa\right)^2.$
Because $\kappa$ and $\tau$ are time-periodic, the elastic energy (the $s$-integral of (\ref{elastic})) is time-periodic, so the time-averaged rate of work done by elastic forces is zero. 

By limiting the size of $\tau$, we confine our attention to motions which are only moderately perturbed from planar motions. If instead we were to allow arbitrary $\tau(s,t)$, we would have essentially arbitrary 3D motions which could involve complex falling and impact dynamics. Such motions are interesting but more challenging to compute accurately, and we do not address them here. With small-to-moderate $\tau$, motions are relatively smooth in time, and $z(s,t)$ is close to time-periodic.  Even with only moderately nonplanar motions it is difficult to consider the full range of possible motions, so we focus on those that are optimally efficient. If $z(s,t)$ is time-periodic (or simply has a finite long-time average), the average rate of work done against both the contact and gravity forces ((\ref{fcontact}) and (\ref{fgravityfriction})) is zero, because both correspond to potential energies that oscillate in time with constant long-time averages. In the simulations, we find that the rates of work done against gravity and contact forces are negligible. The time-averaged power consumption is then the sum of that due to dry friction with the ground and internal viscous dissipation:
\begin{align}
\langle P \rangle = \frac{1}{T}\int_0^T\int_0^L \mathbf{f}_{\mbox{\scriptsize friction}} \cdot \partial_t \mathbf{X} \,ds\, dt + 
\frac{1}{T}\int_0^T\int_0^L \eta_E I \left(\partial_t \kappa\right)^2 \,ds\, dt. \label{AvgPower}
\end{align}
\nn 
We nondimensionalize all quantities (e.g. in equations (\ref{LinAngMom}) and (\ref{AvgPower})) using $T$ as the characteristic time, $L$ as the characteristic length, and $\rho g L $ as the characteristic force. Henceforth the variables are assumed to be dimensionless, but we keep their names the same. 
%, where $\mu_{min} = \min{\mu_f, \mu_b, \mu_n}$. This is the typical magnitude of frictional force on the body. We 
The dimensionless versions of equations (\ref{LinAngMom}) are
\begin{align}
\mbox{Fr} \int_0^1 \partial_{tt} \mathbf{X} ds &= \int_0^1 \mathbf{f}_{\mbox{\scriptsize ext}} ds \quad ; \quad
\mbox{Fr} \int_0^1 \mathbf{X} \times \partial_{tt} \mathbf{X} ds = \int_0^1 \mathbf{X} \times \mathbf{f}_{\mbox{\scriptsize ext}} ds \label{LinAngMomND}
\end{align}
\nn where $\mbox{Fr}$ = $L/gT^2$ has been termed the Froude number \cite{HuNiScSh2009a}. 
As in many previous studies, we assume that $\mbox{Fr}$ is sufficiently small that it can be approximated as zero. Thus we neglect the effect of body inertia, so we do not consider fast motions that involve significant body accelerations \cite{alben2020intermittent}. This simplifies the problem in two ways: it reduces the number of parameters under investigation, and it makes the solutions behave simply under time reparametrization. Now $\mbox{Fr}$ is set to zero, and we can divide the equations (\ref{LinAngMomND}) through by $\mu_f$, so the dependence on friction coefficients is only through the ratios $\mu_n/\mu_f$ and $\mu_b/\mu_f$. Having set $\mbox{Fr}$ to zero, we have eliminated the acceleration terms in (\ref{LinAngMomND}). The result is that body motions are invariant under reparametrization of time, if we assume that $\delta_s$ = 0. In particular, if $\mathbf{X}(s,t)$ is the body motion that corresponds to $\kappa(s,t)$ and $\tau(s,t)$, then $\mathbf{X}(s, ct)$ is the body motion that corresponds to $\kappa(s,ct)$ and $\tau(s,ct)$, for any positive constant $c$. The invariance under time reparametrization was shown in appendix B of \cite{alben2019efficient} and also occurs in many other locomotion models \cite{TaHo2007a,hatton2013geometric,gutman2015symmetries}. Here $\delta_s$ is sufficiently small that the rescaling property holds to a very good approximation. We use the property as follows. We wish to find the kinematics that minimize the average power consumption among all those that achieve a given time-averaged center-of-mass speed $V$. With the rescaling property, any kinematics that give a nonzero average speed can rescaled in time so that the average speed is $V$. 
For a given motion $\mathbf{X}(s,t)$, let the norm of the center-of-mass displacement after a period be $D$ (positive in general). Then for this motion we set $T = D/V$, so it has average speed $V$. The dimensionless average power is (\ref{AvgPower}) divided by $\rho g L^2/T$, but is still denoted $\langle P \rangle$.
We define $\tilde{P}$ to be the dimensionless average power divided by the optimal (i.e. minimal) value for a 2D motion with average speed $V$, which was shown in \cite{alben2019efficient} to be $\rho g \mu_{\mbox{min}}VL$ (dimensional form) or $\mu_{\mbox{min}}D/L$ (dimensionless form), where $\mu_{\mbox{min}} = \mbox{min}(\mu_f, \mu_b, \mu_n)$. This is the power expended by a planar body (i.e. $\|\mathbf{f}_{\mbox{\scriptsize contact}}\| \equiv 1$) sliding uniformly in the direction of minimal friction. Our scaled
average power is
\begin{align}
\tilde{P} \equiv \frac{\langle P \rangle}{\langle P \rangle_{\mbox{opt. 2D}}} = \frac{\langle P \rangle}{\mu_{\mbox{min}} D/L} = \frac{L}{\mu_{\mbox{min}}D}\int_0^1\int_0^1 \mathbf{f}_{\mbox{\scriptsize friction}} \cdot \partial_t \mathbf{X} ds dt + 
c_v \frac{L^2}{D^2}\int_0^1\int_0^1 \left(\partial_t \kappa\right)^2 ds dt. \label{AvgPowerScaled}
\end{align}
\nn where $c_v = \eta_E I V/\rho g L^4\mu_{\mbox{min}}$ is a dimensionless constant that measures the ratio of internal viscous dissipation to external friction. Using the zero-Fr rescaling property, we have reduced the dependence on viscosity and on $V$ to a single parameter $c_v$. We give an order-of-magnitude estimate of $c_v$ for a biological snake (for example, a corn snake \cite{HuNiScSh2009a}). We take $L$ = 30 cm, $g \approx$ 10 m s$^{-2}$, $\rho$ = 1 g cm$^{-3}$, and $\mu_{\mbox{min}}$ = 0.1 \cite{HuSh2012a,wu2020variation}. To estimate $I$, we assume a cylindrical body with thickness $h$ = 3 cm, so $I \approx$ 5 cm$^4$. A typical locomotion speed is $V$ = 10 cm s$^{-1}$. We are not aware of measurements of the effective viscosity of snake tissue (which would only approximate the 
true nonlinear viscoelastic behavior), but an approximation comes from human muscle measurements \cite{schneck1992mechanics}, applied to a similar viscoelastic model of a saithe fish \cite{cheng1998continuous}, with viscosity given as $10^4$--$10^5$ poise. The net result is $c_v \approx$ 10$^{-3}$--$10^{-2}$. 
Smaller viscosities of $10^1$--$10^2$ poise were reported for human muscle and other tissues by \cite{al2018biomechanics}, % p. 14 
resulting in $c_v \approx$ 10$^{-6}$--$10^{-5}$. Therefore, we vary $c_v$ over a wide range, 10$^{-6}$--$10^{-1}$, to consider a wide range of possibilities. 

Next, we describe an algorithm to the determine the kinematics that minimize $\tilde{P}$ among all kinematics that give locomotion at a given speed.

\section{Stochastic optimization \label{sec:Stochastic}}
We now describe the optimization algorithm. We write $\kappa$ and $\tau$ as double Fourier-Chebyshev series:
\begin{align}
\kappa(s,t) &= \sum_{j = 0}^{N_f-1} \sum_{k = 0}^{N_c-1} \left(A_{jk} \cos{2\pi j t} + B_{jk} \sin{2\pi j t}\right)T_k(s) \label{kappaseries}\\
\tau(s,t) &= \left(1 - e^{-(t/t_d)^2}\right) \sum_{j = 0}^{N_f-1} \sum_{k = 0}^{N_c-1} \left(C_{jk} \cos{2\pi j t} + D_{jk} \sin{2\pi j t}\right)T_k(s) \label{tauseries}
\end{align}
\nn where $T_k(s)$ is the Chebyshev polynomial of first kind of degree $k$. We initialize a population of 50 body kinematics, each given by
$2(2N_f-1)N_c$ coefficients $\{A_{jk}, B_{jk}, C_{jk}, D_{jk}\}$ (excluding $B_{0k}$) with $A_{jk}$ and $B_{jk}$ drawn from a Gaussian distribution with standard deviation 
$W_{jk} = 1/j$ if $j \neq 0$, and 1 otherwise.
%\begin{align}
%W_{jk} &= \begin{cases}
%1/j  \;, &\; j \neq 0 \\
%1 \;,  &\; j = 0.
%\end{cases} \label{Wjk}
%\end{align}
\nn The $1/j$ weight corresponds to piecewise continuous functions of time, a minimal type of regularity that we bias the solutions towards because we find empirically that it yields better optima \cite{alben2021efficient}. 
$C_{jk}$ and $D_{jk}$ are drawn from uniform distributions on $\tau_{amp}W_{jk}[-1,1]$, where $\tau_{amp}$ is a nonnegative parameter that limits the size of $\tau$. We compute solutions to (\ref{LinAngMomND}) from $t = 0$ to $t_{\mbox{final}} = 3$, using the second-order BDF method with time step $\Delta t$ = 0.005. At each time step, we solve the six nonlinear equations (\ref{LinAngMomND}) for $\{\mathbf{X}_0(t), \alpha(t), \beta(t), \gamma(t)\}$ using Newton's method with the solution at the previous time step as an initial guess. The integrals in (\ref{LinAngMomND}) are discretized using the trapezoidal rule with grid spacing $\Delta s$ = 0.01. To obtain a good guess for the solution near the initial time, and to decrease the chances of jumping to other branches of solutions at later times, we use a nonzero (but small) Fr value, $10^{-3}$ in (\ref{LinAngMomND}), and solve
the problem as an initial value problem starting from zero tail position/angles and velocity. This provides a good guess (zero tail position/angles) for the solution just after the initial time. Fr is small enough to give a very good approximation to the zero-Fr case, but provides a small amount of inertia that prevents the large accelerations that would occur with jumps to other solution branches. The exponential-in-time factor on the right side of (\ref{tauseries}) causes the body to ramp up from a planar shape over a short time given by $t_d$ = 0.2. This improves convergence during the initial transient in which the body transitions from zero tail velocity to $O(1)$ tail velocity. 

The algorithm runs for a number of generations $N_{gens}$ = 300--600. In each generation, we simulate the motions of each member of the population from $t = 0$ to 3, and compute $\tilde{P}$ from its average over $t \in [2, 3]$, when each body has reached the periodic steady-state motion to a very good approximation. For general kinematics with moderate $\tau$, after a period the body position is the same except for a net translation in the $x$-$y$ plane and a net rotation about the $z$-axis, which can be written $\Delta \alpha = \alpha(3) - \alpha(2)$. Such rotations would lead to the body moving in a large circle over long times, instead of steadily translating across the ground. Here we confine our attention to motions with zero time-averaged rotation. Therefore, we select for motions that have both small $\tilde{P}$ and small $\Delta \alpha$. We assign a ranking 1--50 to each member of population corresponding
to their value of $\tilde{P}$ sorted from low to high (i.e. best to worst). We assign a second ranking 1--50 in the same way but using $|\Delta \alpha|$ instead. Then we define the total ranking to be the maximum of the two rankings. For example, kinematics that give the best, i.e. lowest, $\tilde{P}$, and the worst, i.e. highest, $|\Delta \alpha|$ receive a $\tilde{P}$-ranking of 1, a $|\Delta \alpha|$-ranking of 50, and total
ranking of max(1,50) = 50, the worst possible ranking. We select the half of the kinematics with best total ranking and from each member, form two new members that have the same coefficients $\{A_{jk}, B_{jk}, C_{jk}, D_{jk}\}$ plus small independent random perturbations that are drawn from a uniform distribution on $[-p_{\kappa,jk}, p_{\kappa,jk}]$ for $\{A_{jk}, B_{jk}\}$ and $[-p_{\tau,jk}, p_{\tau,jk}]$ for $\{C_{jk}, D_{jk}\}$: 
\begin{align}
p_{\kappa, jk} &= p_{amp}W_{jk}\left(1-\frac{n}{N_{gens}}\right) \quad ; \quad
p_{\tau, jk} = \tau_{amp}W_{jk} p_{amp}\left(1-\frac{n}{N_{gens}}\right). \label{pert}
\end{align}
\nn Here $p_{amp} = 0.3$ for $p_{\kappa, jk}$, and 0.1 (for half the population) or 0.3 (for the other half) for $p_{\tau, jk}$.  
The maximum perturbation sizes (\ref{pert}) decay to zero as the generation number $n$ increases to $N_{gens}$. Decreasing the perturbation size improves the ability of the algorithm to converge to a local minimum in later generations, similarly to annealing algorithms. After adding the perturbation to $\{C_{jk}, D_{jk}\}$, we scale them if needed so their magnitudes do not exceed $\tau_{amp}W_{jk}$. We thus obtain a new population of 50 at the next generation.

This method selects motions that achieve low $\tilde{P}$ and low $|\Delta \alpha|$ simultaneously. Within the whole space of kinematics, we expect that there are connected subsets or manifolds where $|\Delta \alpha|$ is zero, as in previous work on crawling and swimming with planar kinematics \cite{TaHo2007a,HaBuHoCh2011a,gutman2015symmetries,alben2021efficient}. The algorithm tends to find planar and 3D kinematics that are close to this manifold, with $|\Delta \alpha|$ typically $10^{-4}$--$10^{-2}$.

\begin{table}
\begin{center}
\vspace{1cm}
\begin{tabular}[t]{c|c|c|c|c|c|c|c|c|c|c|c|c|}
\hspace{-0.03cm} $t_{final}$ \hspace{-0.02cm} & \hspace{-0.03cm} $\Delta t$ \hspace{-0.03cm} & \hspace{-0.03cm} $\Delta s$ \hspace{-0.03cm} & \hspace{-0.02cm} $\delta_s$,$\delta_w$, Fr \hspace{-0.02cm} & \hspace{-0.02cm} Pop. Size \hspace{-0.02cm} & \hspace{-0.03cm} $N_{gens}$ \hspace{-0.03cm} & \hspace{-0.03cm} $p_{amp}$ \hspace{-0.03cm} & \hspace{-0.03cm} $N_f$ \hspace{-0.03cm} & \hspace{-0.03cm} $N_c$ \hspace{-0.03cm} & \hspace{-0.02cm} ${\mu_n}/{\mu_f}$ \hspace{-0.02cm} & \hspace{-0.02cm} ${\mu_b}/{\mu_f}$ \hspace{-0.02cm} & \hspace{-0.03cm} $\tau_{amp}$ \hspace{-0.03cm} & \hspace{-0.03cm} $c_v$ \\ \hline
\hspace{-0.03cm}3\hspace{-0.03cm} & \hspace{-0.02cm} 0.005\hspace{-0.015cm} & \hspace{-0.02cm} 0.01\hspace{-0.02cm} & $10^{-3}$ & 50 & \hspace{-0.02cm} 300--600 \hspace{-0.02cm} & 0.1--0.3 & \hspace{-0.02cm} 3--8\hspace{-0.015cm} & \hspace{-0.015cm}3--9\hspace{-0.015cm} & \hspace{-0.015cm}0.1--10\hspace{-0.015cm} & \hspace{-0.02cm}1--5\hspace{-0.015cm} & 0--1 & \hspace{-0.02cm}10$^{-6}$--10$^{-1}$\hspace{-0.02cm} \\ \hline
\end{tabular}
\end{center}
\caption{List of the main parameters and their values or ranges of values used in the present study. \label{params}}
\end{table}

We list the values or ranges of values used here for the most important physical and numerical parameters in table \ref{params}. Next, we present the optimal solutions computed by the model and how they depend on the key parameters: $N_f$, $N_c$, $\mu_n/\mu_f$, $\mu_b/\mu_f$, $\tau_{amp}$, and $c_v$. To keep the presentation somewhat concise, we vary one or two parameters at a time, keeping the others fixed.

\section{Computational results \label{sec:results}}

\subsection{Numbers of modes \label{sec:modes}}

The numbers of Fourier and Chebyshev modes, $(2N_f-1)$ and $N_c$, are important parameters in the optimization. Increasing them increases the range of possible kinematics, allowing for bending and lifting patterns that vary more sharply in space and time. The dimension of the parameter space (for $\kappa$ and $\tau$ together) is $2(2N_f-1)N_c$. It grows rapidly with $N_f$ and $N_c$, which generally increases the number of local optima, and the chance of the population converging to a local optimum that greatly underperforms the global optimum \cite{alben2021efficient}. To study this trade-off in the present problem, we run the optimization with four different nested sets of mode numbers, and two other methods that progressively increase the number of modes during the optimization. 

\begin{figure}
           \begin{center}
           \begin{tabular}{c}
               \includegraphics[width=6.5in]{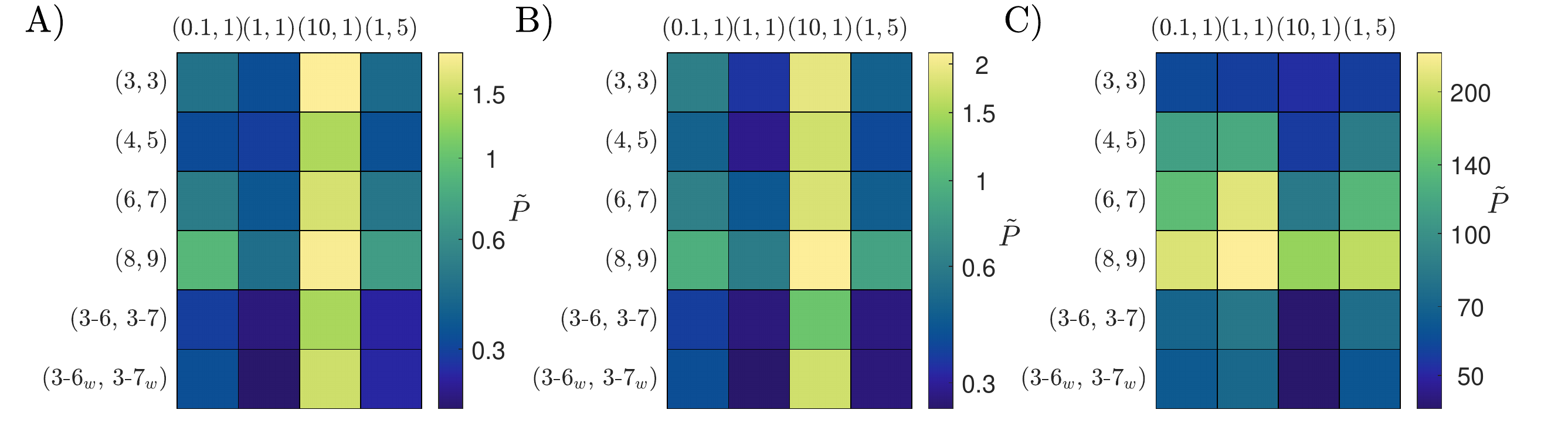} \\
           \end{tabular}
          \caption{\footnotesize Minimum average input power $\tilde P$ achieved by the optimization algorithm for three different viscous damping values, $c_v = 10^{-6}$ (A), $10^{-5}$ (B), and $10^{-1}$ (C). In each panel, six different choices of numbers of modes (labeled at left as an order pair, ($N_f, N_c$)), and four different combinations of friction coefficient ratios (labeled at top as an ordered pair, ($\mu_n/\mu_f, \mu_b/\mu_f$)) are used. The minima are taken over a set of five values, the algorithm with five different random seeds. A different color bar scale is used in each panel to make the intrapanel differences more visible. 
 \label{fig:ModeNumbersFig}}
           \end{center}
         \vspace{-.10in}
        \end{figure}

We fix $\tau_{amp}$ = 0.3 (allowing for small-to-moderate lifting) and run the optimization at four pairs of values of the friction coefficient ratios ($\mu_n/\mu_f, \mu_b/\mu_f$):  (0.1, 1), (1, 1), (10, 1), (1, 5). We compare the
values of $\tilde P$ with six different choices for the mode numbers. In four cases, ($N_f, N_c$) is fixed at (3,3), (4,5), (6,7), 
or (8,9), giving coefficient parameter spaces with dimensions 30, 70, 154, and 270, respectively. In these cases the optimization routine is run for 300 generations, starting from five different sets of random initial coefficients, as described in section \ref{sec:Stochastic}. The 
minimum $\tilde P$ over the five initializations for these mode numbers are shown in the first four rows of each panel of figure \ref{fig:ModeNumbersFig}. Two further cases are considered, in which the numbers of modes start small and increase during the optimization. In one case, ($N_f, N_c$) is (3,3) for the first 200 generations, then is increased to (4, 5) for generations 201-400 (starting from the (3,3) population at generation 200), and finally is increased to (6,7) for generations 401-600 (starting from the (4,5) population at generation 400). The results are shown in the fifth rows of figure \ref{fig:ModeNumbersFig}, labeled (3-6,3-7). In the final case,
the same progressive increase in the numbers of modes is used, but we change the weight function 
$W_{jk}$ to $1/jk$, with $j$ or $k$ replaced by 1 if either is 0.
This alternative weight function damps the coefficients with higher spatial frequencies as well as higher temporal frequencies, so we test whether an added spatial frequency damping is beneficial. The results for this case are shown in the sixth rows of figure \ref{fig:ModeNumbersFig}, labeled (3-6$_w$,3-7$_w$).

Panels A and B of figure \ref{fig:ModeNumbersFig}, corresponding to small damping, show that among the top four rows, i.e. with fixed numbers of modes, (4,5) is always best and (8,9) is usually worst. Optima found with (8,9) modes generally have high-frequency spatial and temporal features, and fit the idea that optimization becomes trapped at local optima that underperform those with fewer modes. The choice (4,5) strikes a balance between having enough modes to represent a variety of motions but not too many high frequency modes and consequent trapping at suboptimal motions. With higher damping (panel C),  (3,3) outperforms the other fixed-mode cases, presumably because the viscous penalization of $\partial_t \kappa$ favors smoother motions. These results for the fixed numbers of modes motivates the two choices in which the numbers of modes increase during the optimization. The idea is to find an optimum with the low number of modes (3,3), and then add the higher modes of the (4,5) and (6,7) cases, hoping that these higher modes will give moderate refinements of the optima with lower numbers of modes, allowing for sharper features but not motions that are dominated by highly oscillatory components. In panels A and B, the choice (3-6$_w$,3-7$_w$) is best except for the friction coefficient pair (10,1), where
(3-6,3-7) is best. These two choices of modes almost always outperform the other cases, except for the first column of panel A where (4,5) is slightly better and panel C, where (3,3) is best in all the columns except the third column. Even in these cases, the
optimal $\tilde{P}$ with (3-6,3-7) and (3-6$_w$,3-7$_w$) are close to those with (3,3) or (4,5). Going forward, we mostly use the
cases (3-6,3-7) and (3-6$_w$,3-7$_w$), with (4,5) in a few cases.

\subsection{Planar optima \label{sec:planar}}

Next, we study how the algorithm performs in the case of no lifting, $\tau_{amp} = 0$, which corresponds to the class of planar motions studied in many previous works \cite{GuMa2008a,HuNiScSh2009a,AlbenSnake2013,alben2021efficient}. The most similar is \cite{AlbenSnake2013}, in which the same modal expansion of curvature was used, but with zero viscous damping only. When $c_v$ is large, the friction term in $\tilde{P}$ becomes insignificant compared to the viscous term, but friction still plays a key role in setting the motion of the body through the dynamical evolution equations. We investigate the planar case first to show the results of the algorithm in this more familiar case. The planar case is also a benchmark against which the benefits of lifting can be assessed. 

\begin{figure} [h]
           \begin{center}
           \begin{tabular}{c}
            \vspace{-0.2in}
             \hspace{-0.2in}
               \includegraphics[width=6.5in]{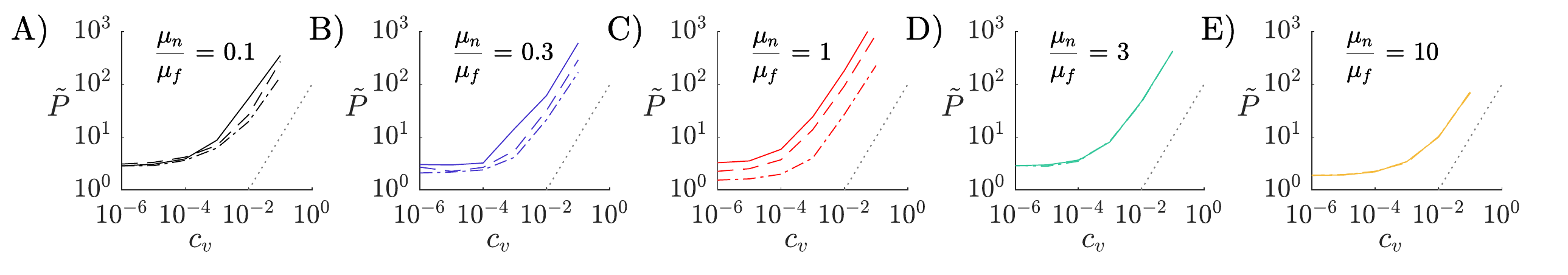} \\
           \end{tabular}
          \caption{\footnotesize The dependence of $\tilde{P}$ on $c_v$ for planar motions ($\tau_{amp} = 0$). Each panel shows data for a different $\mu_n/\mu_f$ value, labeled at the top. Within each panel, three lines are given corresponding to $\mu_b/\mu_f$ = 1 (solid line), 2 (dashed line), and 5 (dashed-dotted line). The dotted line shows the scaling $\tilde{P} \sim c_v$.
 \label{fig:2DcvFig}}
           \end{center}
         \vspace{-.10in}
        \end{figure}

In figure \ref{fig:2DcvFig} we plot $\tilde{P}$ versus $c_v$ for five different $\mu_n/\mu_f$ (increasing from panel A to panel E), and three different $\mu_b/\mu_f$ (solid, dashed, and dashed-dotted lines within each panel). We use $N_f$ = 4 and $N_c$ = 5, which are enough modes to approximate a wide range of motions including those in \cite{AlbenSnake2013}. The curves are approximately flat near $c_v = 10^{-6}$, which is small enough that frictional power dissipation dominates that from viscosity, and the optima are essentially unchanged with further decreases of $c_v$. At $c_v = 10^{-1}$, the other end of the range,
$\tilde{P}$ grows approximately linearly with $c_v$. Here the optimal motions have reached their upper asymptotic limit, and are essentially unchanged with further increases of $c_v$. The second term in the rightmost expression in (\ref{AvgPowerScaled}) is dominant, and scales as $c_v$ when $\kappa$ remains fixed and $c_v$ increases. In panels D and E,
$\mu_n/\mu_f$ = 3 and 10, there is little dependence on $\mu_b/\mu_f$, because the optimal solutions involve forward motion only. The dependence on $\mu_b/\mu_f$ is largest in panel C ($\mu_n/\mu_f$ = 1), and decreases at smaller $\mu_n/\mu_f$ (panels B and A). In almost all cases, larger $\mu_b/\mu_f$ is better (decreases $\tilde{P}$).

\begin{figure}
           \begin{center}
           \begin{tabular}{c}
\hspace{-0.2in}
               \includegraphics[width=5.7in]{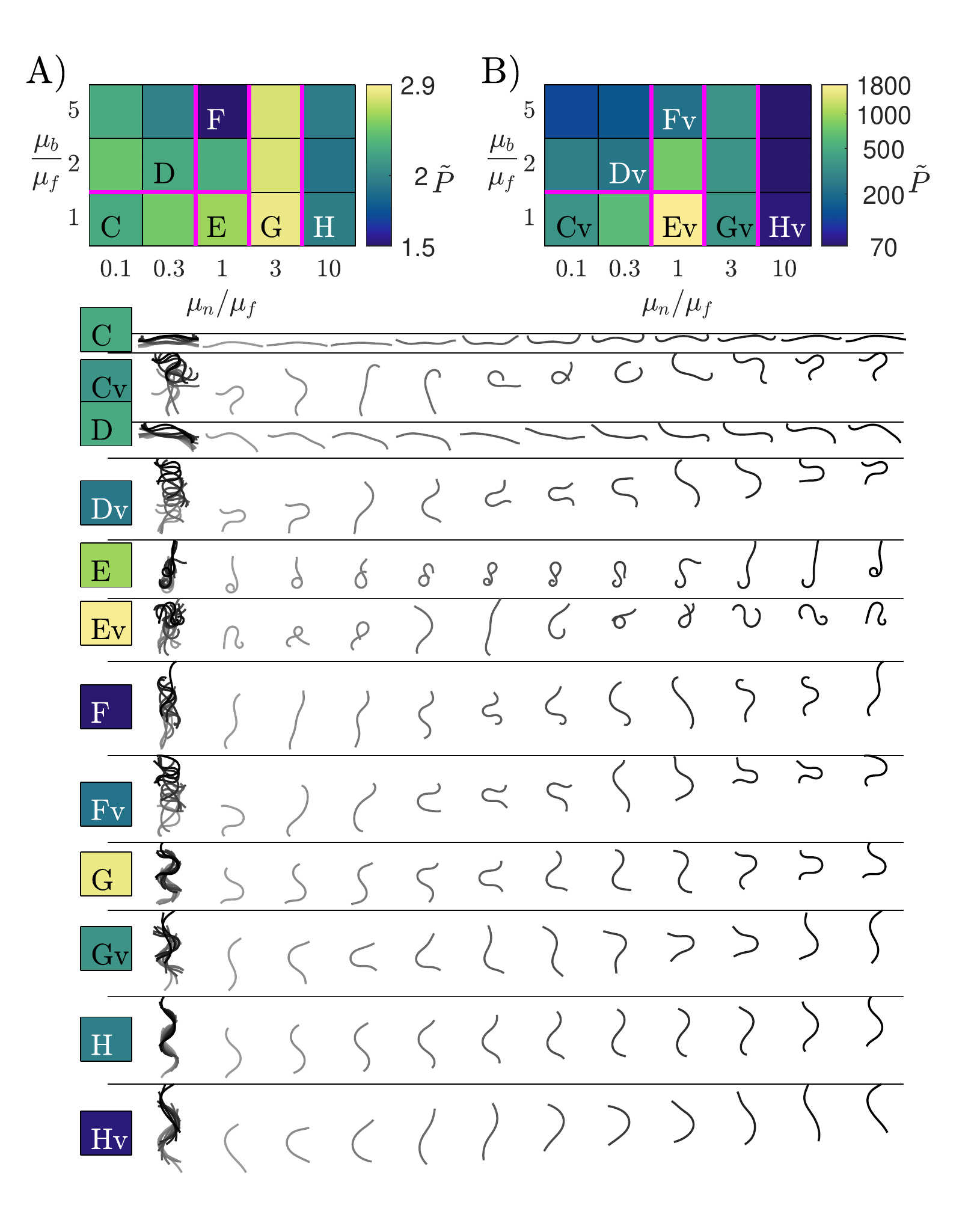} \\
           \end{tabular}
          \caption{\footnotesize Values of input power $\tilde{P}$ for optimal planar motions with damping constant $c_v$ = 0 (A) and 0.1 (B), across a 5-by-3 grid of values of the friction coefficient ratios $\mu_n/\mu_f$ and $\mu_b/\mu_f$. In panels A and B the friction coefficient space is divided by magenta lines into 6 regions, where similar types of optimal motions occur. At the points C--H (panel A) and Cv-Hv (panel B), examples of these motions are shown by the sequences of snapshots in the lower portion of the figure. 
 \label{fig:BodySnapshots2DFig}}
           \end{center}
         \vspace{-.10in}
        \end{figure}

Typical optimal planar motions are shown in figure \ref{fig:BodySnapshots2DFig}, computed with the mode choice (3-6,3-7) and five random initial sets of coefficients.
The minimal values of $\tilde{P}$ are plotted in friction coefficient space in panels A and B, with zero and large viscous damping respectively ($c_v$ = 0 and 0.1). The magenta lines serve to divide friction coefficient space into six regions, each characterized by a typical motion. In the two rightmost columns, $\mu_n/\mu_f$ = 3 and 10, the optimal motions are essentially independent of  
$\mu_b/\mu_f$, as already noted for the values of $\tilde{P}$ in figure \ref{fig:2DcvFig}. The motions are retrograde traveling motions, similar to those found using a Newton-based rather than population-based optimization approach in \cite{AlbenSnake2013} (without viscous damping). For the box labeled H at $\mu_n/\mu_f$ = 10 in panel A, the body motion is shown next to the letter H near the bottom of the figure, as a set of 11 snapshots over a period of motion.
The snapshots are uniformly rotated so that the (arbitrary) direction of time-averaged center-of-mass displacement is up the page. 
The snapshots are shown twice, first in their true physical positions immediately adjacent to the letter H, as an overlapping set of shapes that vary from light gray to black as time increases. Then, extending rightward, the snapshots are spread out across the page with
a fictitious horizontal displacement that makes it possible to see each snapshot individually (the true horizontal displacement over a period is zero). The overlapping snapshots show that the body moves along an undulatory path, and the spread-out snapshots show that a traveling wave of curvature moves from head to tail, against the direction of locomotion. At $\mu_n/\mu_f$ = 3, labeled G, the motion is again a retrograde traveling wave, but the body is more compressed vertically, with larger curvature. The spread-out snapshots show that the curvature magnitude varies more in time, unlike for motion H. Essentially the same motion was observed in \cite{AlbenSnake2013}, and it was noted that the increase in curvature in G relative to H increases the normal component of motion, which counteracts the decrease in $\mu_n/\mu_f$ in G relative to H. Here we find 
$\tilde{P}$ increases as $\mu_n/\mu_f$ decreases from 10 to 3, also consistent with the results in \cite{AlbenSnake2013} where
the efficiency, the reciprocal of $\tilde{P}$ with zero $c_v$, was used instead. With large viscous damping
the corresponding motions are labeled Gv and Hv. Again,  $\tilde{P}$ and the optimal motions are independent of 
$\mu_b/\mu_f$ in this region. Compared to G and H, Gv and Hv have somewhat larger deflection amplitudes and smaller curvatures. Smaller curvature clearly decreases the integrand in the second term on the right side of (\ref{AvgPowerScaled}).
Larger deflections increase $D$, the displacement within a period, which decreases the viscous damping term in $\tilde{P}$ more strongly than the friction term, the first term on the right side of (\ref{AvgPowerScaled}). Hence larger displacements over a period are preferred more strongly when the viscous damping constant $c_v$ is large.

For $\mu_n/\mu_f$ = 1 and smaller, there are different types of optimal motions for $\mu_b/\mu_f$ = 1 and $\mu_b/\mu_f > 1$. In the isotropic case (E), the optimal motion can be considered as a type of concertina motion. In the first three snapshots, the rear portion of the body curls up, thus moving forward, while the front portion, the ``anchor," is approximately fixed. In snapshots 4-7, the rear curl shrinks, and the part of the body in the curl moves tangentially into a front curl. In snapshots 8-11, the front curl opens, swinging this part of the body forward, and then the rear of the body curls up, also moving forward. In general, either the front or rear portion, the ``anchor," is approximately fixed while the rest of the body moves forward.  The isotropic case was studied extensively in \cite{alben2019efficient}, and different types of concertina motions were found computationally and analytically. Motion Ev, the more viscous optimum, involves larger swinging motions and somewhat less curling, as well as more self-intersecting (which is not prevented for simplicity, as the self-intersection is not of primary importance in this work). As before, the curvatures are smaller while the net displacement is larger in the viscous case. The isotropic case has the largest $\tilde{P}$ in the viscous case (panel B), while it is somewhat smaller than the $\mu_n/\mu_f$ = 3 case without viscosity (panel A). Motions F and Fv exemplify the region $\mu_n/\mu_f$ = 1 and $\mu_b/\mu_f > 1$. The body snapshots in both cases consist of a repeated flexing and unflexing, together with a slightly forward-propagating curvature wave. A similar but more fore-aft symmetric motion was found in this regime in \cite{AlbenSnake2013}. For $\mu_n/\mu_f  <  1$, the optimal motions with zero viscosity (C and D) involve small amplitude motions about a position that is mostly perpendicular to the direction
of locomotion, and hence experiences low drag. These motions also have a small displacement per period, which is unfavorable at large viscosity. There the motions (Cv and Dv) have much larger degrees of flexing with a swinging motion (Cv) or a forward-propagating wave of curvature (Dv). 

\subsection{Lifting amplitude \label{sec:lifting}}

\begin{figure} 
           \begin{center}
           \begin{tabular}{c}
\hspace{-0.5in}
               \includegraphics[width=6in]{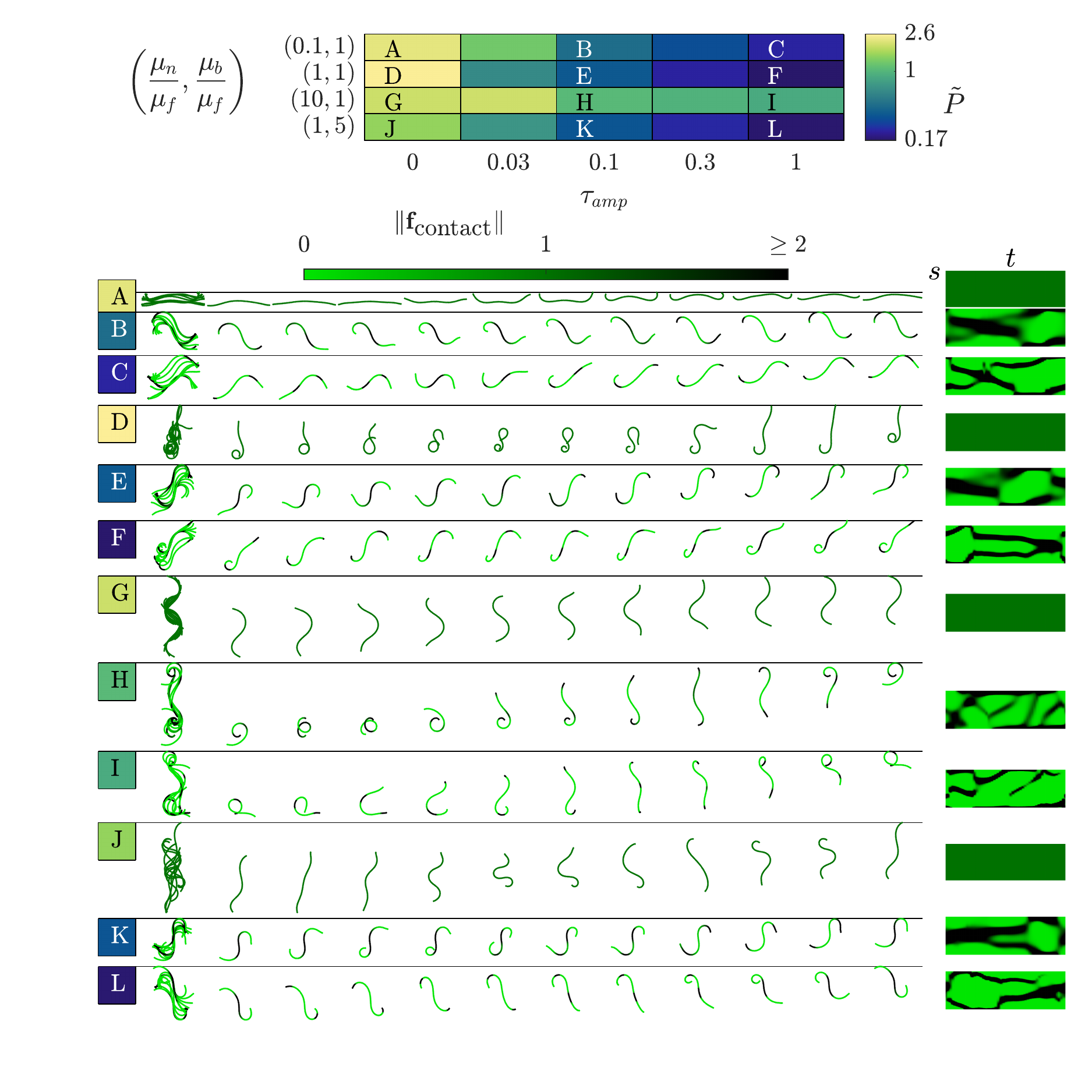} \\
           \end{tabular}
          \caption{\footnotesize 
For zero internal viscosity ($c_v = 0$), minimum $\tilde{P}$ and corresponding motions for different lifting amplitudes $\tau_{amp}$ at four different friction coefficient ratio pairs. At the top is a color plot showing the minimum $\tilde{P}$ found at each
$\tau_{amp}$ (horizontal axis) and friction coefficient ratio pair (vertical axis). For the choices $\tau_{amp}$ = 0, 0.1, and 1,
snapshots the corresponding optimal motions are shown below (labeled A-L), with snapshots displaced horizontally. To the right of each row of snapshots is a color plot showing the ground contact force density $\left\lVert\mathbf{f}_{\mbox{contact}}\right\rVert$ versus
%arc length $s$ and time $t$, over one period of time.
 \label{fig:BodySnapshotsVsTauDamping0Fig}}
           \end{center}
         \vspace{-.10in}
        \end{figure}

We now consider how the optimal motions change when the body is allowed to lift off of the plane, i.e. when $\tau_{amp}$ is increased from zero. To limit the number of cases under discussion, we focus on just four choices of friction coefficient pairs 
($\mu_n/\mu_f, \mu_b/\mu_f$):  (0.1, 1), (1, 1), (10, 1), (1, 5), listed along the left boundary of the color plot at the top
of figure \ref{fig:BodySnapshotsVsTauDamping0Fig}. Along the bottom boundary, $\tau_{amp}$ varies from 0 to 1. The viscous damping constant $c_v$ is set to 0. Moving from left to right in the color plot, we see that the optimal $\tilde{P}$ drops with increasing $\tau_{amp}$ (except for the very slight increase just to the right of point G). It is reasonable that the optimal $\tilde{P}$ decreases, because the set of possible motions at a given $\tau_{amp}$ includes those at smaller $\tau_{amp}$. %In some cases (such as one with $c_v = 0.1$ in the next figure) 
Enlarging the set of possible states could cause the algorithm to find a worse local optimum, but not here. To show how the optimal motions change with lifting, we use A-L to label the subsets of cases with $\tau_{amp}$ = 0, 0.1, and 1 at each of the four friction pairs, and show the body snapshots below. 

Moving from the planar motion A to the nonplanar motion B, the type of motion changes completely, from a small amplitude deflection about a straight body, to a ``walking" type of motion. The snapshots are colored according to the contact force magnitude (green-black scale above the snapshots), where lifted regions of the body are in green, and those strongly contacting the ground are in black. The color maps at the right show, for each set of snapshots, the distribution of contact forces in the space of arc length $s$ and time $t$, over one period. Motion B involves small perturbations about an s-shaped curve. The contact map at right shows that motion B involves two phases. The first phase is shown by two crossing diagonal black bands. Thus two contact regions move from the ends to the middle and back to the ends. In this phase, the snapshots show that the ends are lifted and moved in the forward direction. In the second phase, the black regions are confined to the ends. Here the middle part of the body is lifted and moved in the forward direction. From A to B, $\tilde{P}$ is reduced by about a factor of five. Motion C, with $\tau_{amp}$ increased to 1, is similar to B, though the bands of lifting are sharper and $\tilde{P}$ is reduced by an additional factor of almost two. In principle, alternately lifting parts of the body and moving them only when they are off the ground can reduce the power done against friction to very small values. Motions D, E, and F show the sequence of optima with increasing $\tau_{amp}$ when friction is isotropic.  The snapshots and contact maps of E and F are fairly similar to those of B and C. The mean body configurations in E and F are more aligned with direction of motion than B and C, probably because of the increase in $\mu_n/\mu_f$. The mean power $\tilde{P}$ is reduced by about 25\% from B/C to E/F, perhaps due to slight differences in the relative amounts of tangential and normal motions.%, as well as the greater displacement in E/F. 

Motions G, H, and I correspond to $\mu_n/\mu_f$ = 10 and $\mu_b/\mu_f$ = 1. The planar motion (G) is a basic lateral undulation, a retrograde traveling wave. The lifted motions (H and I) have snapshots that start with the body curled up. Then part of the body straightens and moves forward. It is mostly lifted off the ground except at one end, its more forward point, that is in contact with the ground, and slides forward along the ground. In the last few snapshots (in H and I), the forward end curls up, drawing the rest of the body forward behind it. The overall motion is similar to a concertina motion, with an anchor formed by the curled region, and the straight region extended or retracted forward, mostly lifted, except at one end. From G to H to I, $\tilde{P}$ drops by 46\% and then by 17\%, smaller reductions from lifting than at the other friction pairs. Motions H and I combine lifting with sliding mainly in the tangential direction, in contrast to B/C and E/F, where the body is oriented transverse to the direction of motion, so the normal friction coefficient applies more strongly for the sliding that occurs.

Motions J, K, and L apply with $\mu_n/\mu_f$ = 1 and $\mu_b/\mu_f$ = 5. K and L are very similar to E and F, both in the snapshots' shapes and in the values of $\tilde{P}$ (those for K/L are within a few percent of those for E/F). There is little backward motion, so the difference in $\mu_b/\mu_f$ is not very important.

In some cases (D, H, and I) there is apparently self-intersection at certain times. This partly due to the 2D projection of the images, and in any case is not a major concern because with slight modifications the motions avoid self-intersection.

\begin{figure}
           \begin{center}
           \begin{tabular}{c}
\hspace{-0.5in}
               \includegraphics[width=6in]{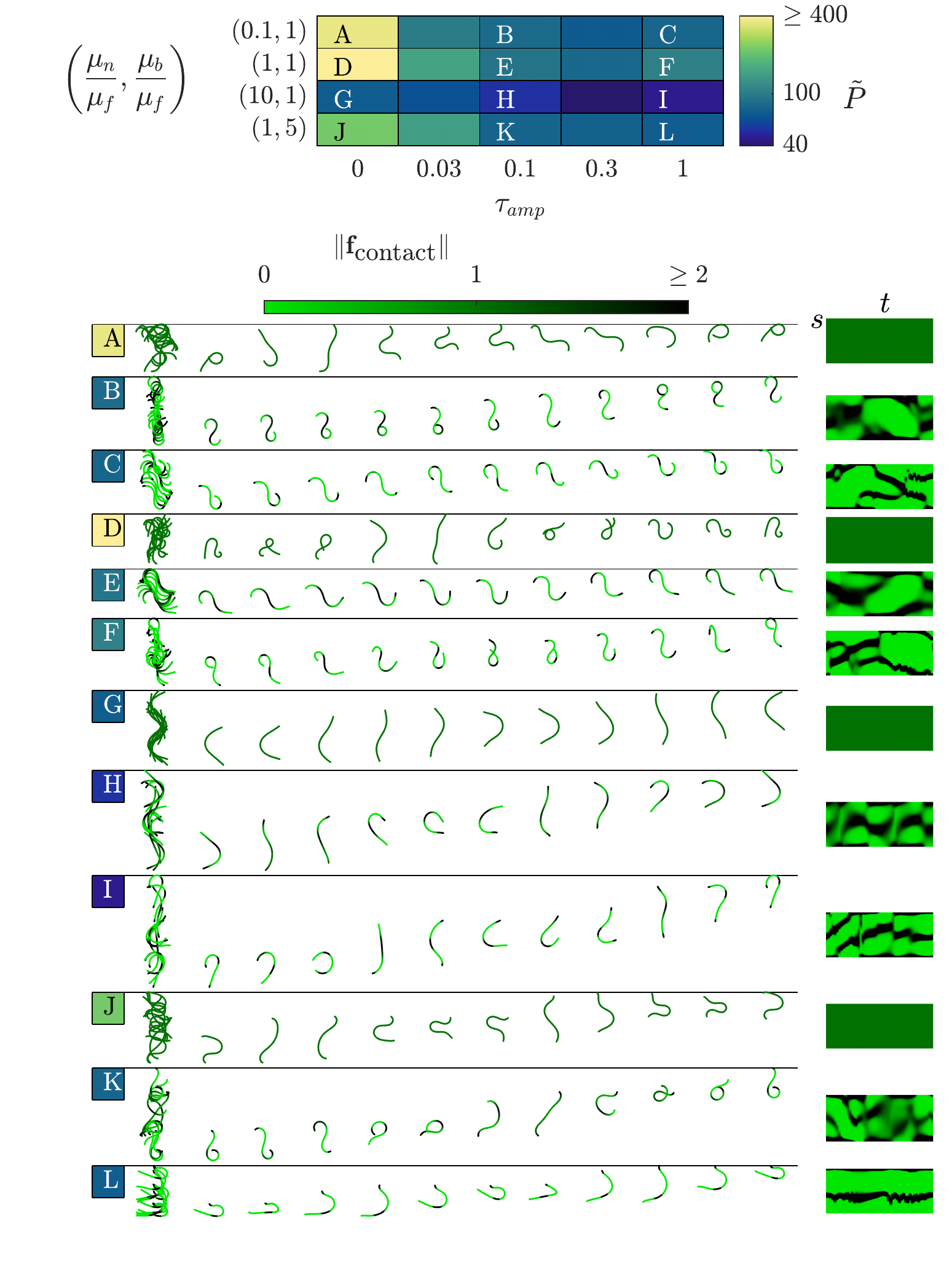} \\
           \end{tabular}
          \caption{\footnotesize For a relatively large internal viscosity  ($c_v = 0.1$), the same quantities as in figure \ref{fig:BodySnapshotsVsTauDamping0Fig}.
 \label{fig:BodySnapshotsVsTauDamping01Fig}}
           \end{center}
         \vspace{-.10in}
        \end{figure}

In figure \ref{fig:BodySnapshotsVsTauDamping01Fig} we consider the same quantities but instead of zero viscosity, we have large viscosity ($c_v = 0.1$). As with zero $c_v$, there is a large change in motions and $\tilde{P}$ values as $\tau_{amp}$ increases from 0 to 0.03 or 0.1, and a smaller change as $\tau_{amp}$ increases further to 1. This is particularly true for the friction
coefficient pairs (0.1,1) and (1,1), and somewhat less so at (10,1) and (1,5). As in figure \ref{fig:BodySnapshotsVsTauDamping0Fig}, motions B and C mainly involve an alternating pattern of lifting and moving of the outer regions and of the central region. Here, however, the body curves more and obtains a larger displacement over one period.
The main advantage probably is to increase $D$ in the $c_v$-term in (\ref{AvgPowerScaled}), thereby decreasing $\tilde{P}$. Motions E and F are fairly similar, though interestingly E resembles C more than B, and F resembles B more than C, despite the different $\tau_{amp}$ values. Motion H qualitatively resembles G even though G has no lifting. H has an alternating 2-2 contact pattern as it flexes symmetrically to the left and to the right. The main contacts are at the head and midbody when the tail swings forward, and at the tail and midbody when the head swings forward. $\tilde{P}$ is reduced by about 30\% from G to H. Motion I is similar to H but less symmetrical, and gives a further 14\% reduction in $\tilde{P}$. With lifting, the friction coefficient pair (10,1) has the highest power with zero $c_v$ (HI at the top figure \ref{fig:BodySnapshotsVsTauDamping0Fig}) but the lowest power with large $c_v$ (figure \ref{fig:BodySnapshotsVsTauDamping01Fig}). Motions K and L are very different from each other, but have about the same $\tilde{P}$, about a factor of three less than that of the nonlifting optimum J. K involves large bending to the left and right, with some similarities to H and I. L has bending to one side only, with two almost-fixed contact regions near the head and the midbody. It resembles J in that it is a repeated bending and unbending motion, but to one side only. A combination of backward and normal friction acts at each contact region at different times, pushing the body forward. 

For the first three of the friction coefficient pairs, the smallest $\tilde{P}$ is obtained at $\tau_{amp}$ = 0.3, and for the fourth,
the smallest value occurs at $\tau_{amp}$ = 1, but it is within 7\% of the values at $\tau_{amp}$ = 0.1 and 0.3. Because friction is negligible in $\tilde{P}$ at large $c_v$, a large degree of lifting is less important in terms of avoiding frictional power dissipation. But a moderate degree of lifting is still much better than small lifting, because it changes the correspondence between body bending and locomotion. It allows for a smaller viscoelastic (bending) dissipation for a given average speed of locomotion. 

To summarize, at both zero viscous damping and large viscous damping there are large changes in the optimal motions and large improvements in $\tilde{P}$ when $\tau_{amp}$ increases from 0 to 1. Most of the changes occur in the increase from 0 to 0.1, with smaller changes over the increase from 0.1 to 1. The optimal motions in the latter range are characteristic of the regime of moderate lifting, in which the body is mostly extended in the $x$-$y$ plane with a $z$-extent that is much smaller, but large enough to lift completely off the ground at most points. In some cases, a lower $\tilde{P}$ is obtained at 
$\tau_{amp}$ = 0.3 than at 1. In most cases, the optimization routine is faster and more robust at $\tau_{amp}$ = 0.3 than at 1. At $\tau_{amp}$ = 1, the lifting may be large enough that the body tips over due to gravity, and a tumbling motion occurs which is not resolved by the time-stepping algorithm. In order to avoid such cases while examining the effects of the friction coefficients in more detail, we therefore fix $\tau_{amp}$ at 0.3, which gives a good representation of motions with a moderate amount of lifting.

\subsection{Effects of frictional anisotropies and damping \label{sec:anisotropies}}

\begin{figure} [h]
           \begin{center}
           \begin{tabular}{c}
\hspace{-0.2in}
               \includegraphics[width=6in]{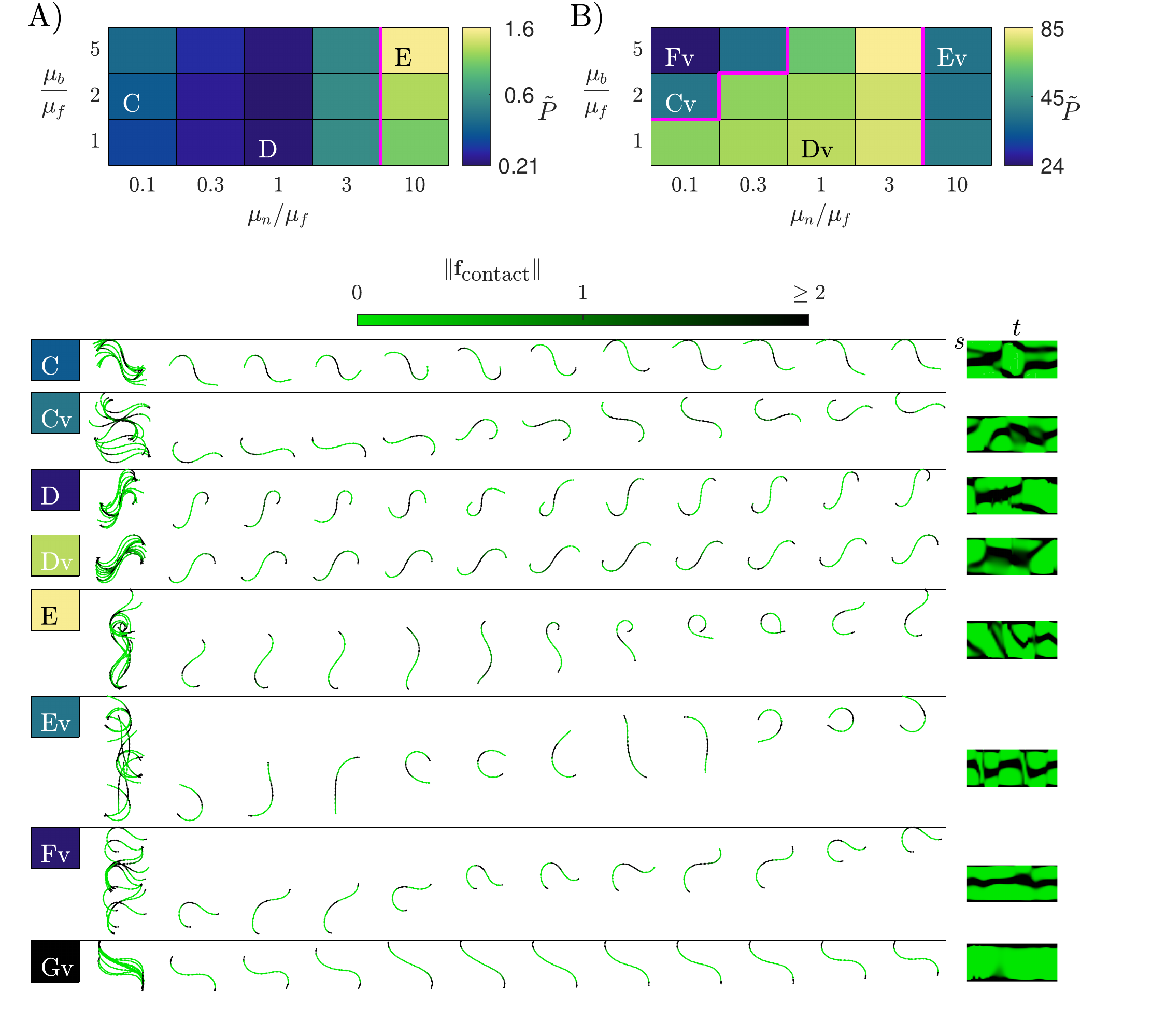} \\
           \end{tabular}
          \caption{\footnotesize  Values of input power $\tilde{P}$ for optimal motions with lifting amplitude $\tau_{amp}$ = 0.3 and damping constants $c_v$ = 0 (A) and 0.1 (B), across a 5-by-3 grid of values of the friction coefficient ratios $\mu_n/\mu_f$ and $\mu_b/\mu_f$. In panels A and B the friction coefficient space is divided by magenta lines into different regions where similar types of optimal motions occur. At the points C--E (panel A) and Cv-Fv (panel B), examples of these motions are shown by the sequences of snapshots in the lower portion of the figure. 
 \label{fig:BodySnapshots3DVsDampingAndFrictionFig}}
           \end{center}
         \vspace{-.10in}
        \end{figure}

We now consider the optimal motions and $\tilde{P}$ values across the 5-by-3 grid of friction coefficient ratios. We use ten random initializations with each of the (3-6,3-7) and (3-6$_w$,3-7$_w$) mode choices. We fix $\tau_{amp}$ at 0.3, and vary $c_v$ from 0 to 0.1 (large). In figure \ref{fig:BodySnapshots3DVsDampingAndFrictionFig} we plot the optima at the two extreme $c_v$ only, 0 in panel A and 0.1 in panel B. We have also computed results at intermediate values, and we find that around $c_v = 10^{-3}$ they smoothly transition between the behaviors at the extremes. With zero $c_v$ (panel A), there are essentially two types of optima, one for $\mu_n/\mu_f \leq 3$ and another at the largest $\mu_n/\mu_f$ = 10. The two regimes are divided by the magenta line in panel A. For $\mu_n/\mu_f \leq 3$ and all $\mu_b/\mu_f$, the optimal solution is an s-shaped body with alternate lifting of the body ends and the body middle, similar to those shown in figure \ref{fig:BodySnapshotsVsTauDamping0Fig}B, C, E, F, K, and L, with modest changes depending on the friction coefficient ratios. Examples are shown by
the motions marked C and D in figure \ref{fig:BodySnapshots3DVsDampingAndFrictionFig}. The values of $\tilde{P}$ are almost independent of  $\mu_b/\mu_f$ at $\mu_n/\mu_f$ = 1 and 3, but have a noticeable variation at $\mu_n/\mu_f$ = 0.1 and 0.3.
At $\mu_n/\mu_f$ = 10, the optimum (E) is a combination of undulation and curling/uncurling, together with lifting, as in figure
 \ref{fig:BodySnapshotsVsTauDamping0Fig}H and I. 

With large viscosity, $c_v$ = 0.1 (panel B), the friction coefficient space divides into three types of optimal motions rather than two. At 
$\mu_n/\mu_f$ = 1 and 3 the alternating lifting motion (shown by Dv) is again optimal. At $\mu_n/\mu_f$ = 10, a symmetric bending motion with alternating pairs of contacts (Ev) is optimal, as in figure \ref{fig:BodySnapshotsVsTauDamping01Fig}H and I. At small  $\mu_n/\mu_f$ and $\mu_b/\mu_f > 1$, different optima are seen, shown by Cv and Fv. The first four snapshots of Cv show the middle part of the body lifted and placed forward. Snapshots 4-7 show curling and uncurling of the right side of the body, which pushes the left side of the body forward. Snapshots 8-11 show curling and uncurling of the left side of the body, which pushes the right side of the body forward. 
Motion Fv is a similar pattern of curling and uncurling, but by the left side of the body only. Curling pushes the left side forward, and uncurling pushes the right side forward. Meanwhile, there are three almost fixed contact regions at the two ends and the middle of the body, which resembles the contact pattern at certain times in Cv, but not others.

At the bottom of the figure is an example of a different type of optimal motion, labeled Gv. The friction coefficients are the same as for E and Ev ($\mu_n/\mu_f$ = 10, $\mu_b/\mu_f$ = 5), but $c_v = 10^{-4}$, intermediate between the values in panels A and B. The contacts occur exclusively at the two ends throughout the motion. Because these are oriented in the tangential direction, frictional dissipation occurs mainly with the minimal drag coefficient $\mu_f$. There is also backward movement at the contacts, but it is much smaller than the forward movement because $\mu_b/\mu_f$ = 5.

\subsection{Optima with traveling-wave contacts \label{sec:sidewinding}}

\begin{figure} [h]
           \begin{center}
           \begin{tabular}{c}
              \hspace{-0.2in}\includegraphics[width=6.5in]{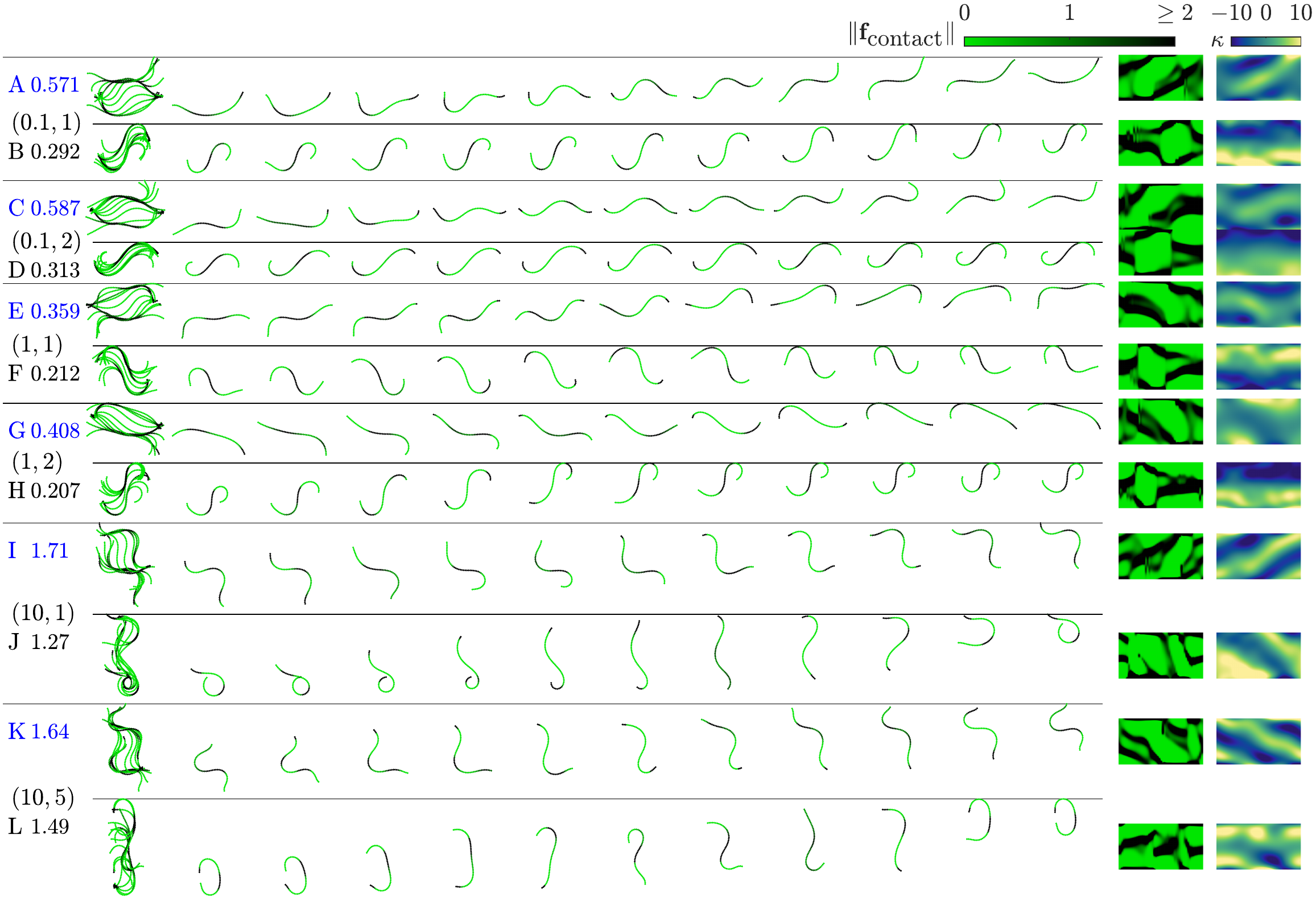} \\
           \end{tabular}
          \caption{\footnotesize Six examples of local optima with traveling-wave contact force distributions (labeled in blue at left, A, C, E, G, I, and K), paired with the global optimum (B, D, F, H, J, and L respectively) found at that set of friction coefficient ratios, listed in parentheses at the left. Also listed at left are the $\tilde{P}$ values. 
$c_v$ is $10^{-6}$ in B, F, H, J, K, and L, and 0 in the other six cases. To the right of each row of snapshots are plots showing the contact force distributions and curvature distributions versus $s$ and $t$. 
 \label{fig:MakeSidewindingComparisonFig}}
           \end{center}
         \vspace{-.10in}
        \end{figure}

One of the most interesting types of lifting motions in biological snakes is the sidewinding motion, which consists of a traveling wave of curvature synchronized with a traveling wave of lifting, with a phase difference between the two \cite{gray1946mechanism,jayne1986kinematics,marvi2014sidewinding,rieser2021functional}. Interestingly, for some random initializations our optimization algorithm converges to these types of motions, across friction coefficient space. This only occurs for low viscous damping, $c_v = 0$ or $10^{-6}$ usually.
So far in this paper we have presented only the best local optimum we have found at a given parameter set, and these are never the sidewinding-type motions. We now show, in figure \ref{fig:MakeSidewindingComparisonFig}, examples of the sidewinding-type local optima. Six different examples are shown, labeled A, C, E, G, I, and K in blue at left. Each case has a different pair of friction coefficient ratios, in parentheses below and to the right of the blue letter label. Below each of the six cases is the best local optimum found at the same friction coefficient ratios, labeled B, D, F, H, J, and L at left. These are alternating lifting motions (B, D, F, H) or curling and sliding motions (J and L) that are similar to those already discussed. Next to each letter label is the value of $\tilde{P}$ for that motion, and the values for the sidewinding motions vary from slightly higher (K versus L) to about a factor of two higher (A versus B, C versus D, etc.) than those of the best local optima. 

To the right of each set of snapshots are the contact force maps and curvature maps in $s$-$t$ space.
For the sidewinding motions (A, C, E, G, I, and K), both maps have diagonal bands with about the same slopes, showing unidirectional traveling waves of contact force and curvature that are approximately synchronized. The optimal motions (B, D, F, H, J, and L) do not show unidirectional traveling waves except for motion J, and there the diagonal bands in the contact and curvature maps have different slopes, corresponding to waves moving at different speeds. Another distinctive feature of the sidewinding motions can be seen by examining the sets of snapshots in physical space, just to the right of the $\tilde{P}$ values. The black regions (where the body contacts the ground) of all the snapshots together 
trace out approximately continuous line segments on the ground. Between the black regions are almost-parallel arrays of green regions, lifted parts of the snake body that are being moved from one black region to the next. When the motion is repeated over multiple periods, the sets of black line segments form a series of parallel tracks on the ground, as for biological sidewinding snakes \cite{gray1946mechanism,jayne1986kinematics}. 
In A, C, E, G, I, and K, the tracks are almost orthogonal to the direction of locomotion (up the page), whereas biological tracks are typically at an oblique angle \cite{gray1946mechanism,jayne1986kinematics,lillywhite2014snakes}. In figure \ref{fig:MakeSidewindingComparisonFig} the sidewinding motions separate into two groups: $\mu_n/\mu_f \leq 1$ (ACEG), where the lifted (green) body segments are at an oblique angle to the tracks and to the direction of locomotion, and
$\mu_n/\mu_f = 10$, where the lifted segments are perpendicular to the tracks and parallel to the direction of locomotion. Here the tracks are shorter and the body has a larger displacement per period.

\section{Simple theoretical motions \label{sec:theoretical}}

\begin{figure} [h]
           \begin{center}
           \begin{tabular}{c}
               \includegraphics[width=6.5in]{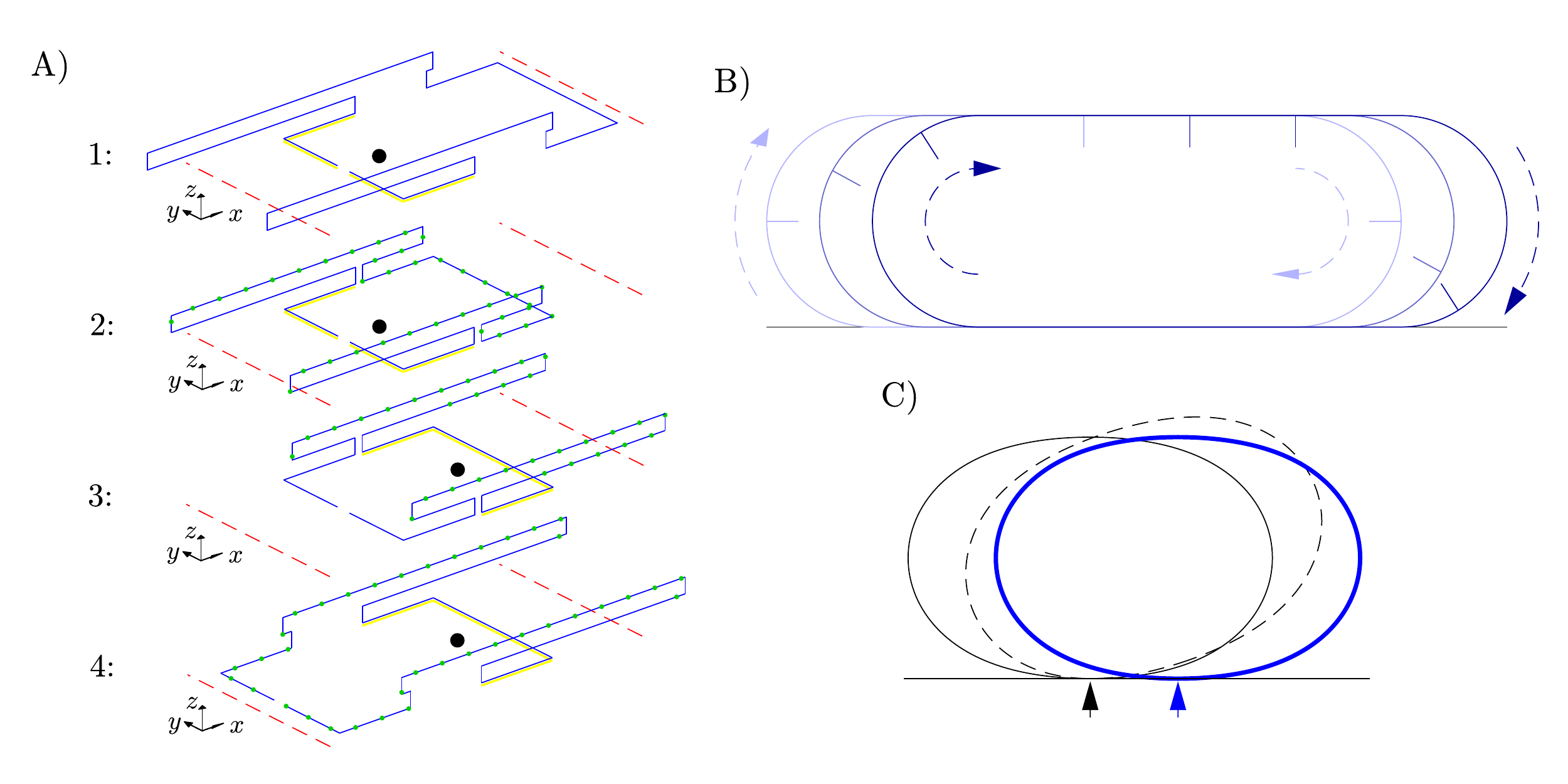} \\
           \end{tabular}
          \caption{\footnotesize Examples of motions with very little work done against friction. A) Walking motion, shown by four snapshots from top to bottom. B) Tank-treading rolling motion, show by a sequence of three snapshots (light, medium, and dark blue). C) Falling-center-of-mass rolling motion, shown by a sequence of three snapshots (solid black line, dashed black line, and blue line). 
 \label{fig:WalkingRollingFig}}
           \end{center}
         \vspace{-.10in}
        \end{figure}

In figure \ref{fig:WalkingRollingFig} we show theoretical motions that use lifting to locomote with essentially zero work done against
friction. There is still viscous dissipation for these motions, but it can be made small potentially by certain versions of these motions. Panel A shows a sequence of four snapshots of a walking type of motion, similar in principle to the optima computed for $\mu_n/\mu_f \lesssim 1$.
In each snapshot of panel A, the body is bent into a shape with the $x$-$z$ plane as a plane of symmetry, in order to extend its base of support along the $y$ direction, and make it stable to small perturbations that involve a rotation about the $x$ axis. The body has regions in contact with the ground---the ``feet"---and an elevated region that is used to shift the center of mass from one foot to the other. Two dashed red lines are used as fixed guides to show how far the body locomotes. In snapshot 1, the left foot, highlighted in yellow, is in contact with the ground, and the body's center of mass, a black dot, lies within its convex hull, so the body is stable. From snapshot 1 to 2, the right foot is lifted slightly, then brought closer to the left foot, and meanwhile the elevated portion shifts rightward to keep the center of mass fixed. Small green dots in snapshots 2--4 show the regions of the body that have moved from the previous snapshot. From snapshot 2 to 3, the body puts both feet on the ground, then shifts the elevated region so the center of mass lies over the right foot, then lifts the left foot. From snapshot 3 to 4, the left foot is moved leftward, and simultaneously the elevated region shifts rightward to keep the center of mass fixed. Now the body can return from snapshot
4 to snapshot 1 (but translated leftward), shifting the elevated region and the center of mass leftward, and repeat the process. As with the computed optima, the body is a single continuous segment here. The computed optima, which involved lifting the middle portion and ends of an s-shaped body, are a smoother type of walking motion, which have smaller viscous dissipation than in panel A (or even a slightly regularized version that removes the sharp corners). With a finite number of modes, the computed optima probably cannot have a frictional dissipation that is precisely zero, as it is for the motion in panel A.

In panels B and C we examine two rolling types of motions, described as kinematic and dynamic rolling
in \cite{sastra2009dynamic}. A vast number of shape-changing rolling robots have been created using these types of motions, e.g. \cite{armour2006rolling,puopolo2016velocity,li2021electrically,shah2021soft}. Panel B shows three successive snapshots of a body performing a tank-treading motion. Portions of the body at the left are rolled off the ground into the semicircle at the left end while portions at the right are rolled out of the semicircle at the right end, onto the ground.  The dashed lines with arrows show the directions of rolling, and the hash marks show how fixed material points move between the three snapshots. In the tank-treading motion, the region in contact with the ground is stabilized by static friction, which is approximated by the sliding friction model
with the small $\delta_s$ term we have used (equation (\ref{fgravityfriction})) as described in \cite{alben2019efficient}. The tank-treading motion is called kinematic rolling, because the body is always at a stable equilibrium and inertia plays no role. In panel C, dynamic rolling, the body
starts in a stable elliptical shape (solid black line), and then changes its shape to the dashed line. If friction is sufficiently large, the ground contact remains approximately fixed. The dashed line shape is gravitationally unstable, so the body rolls rightward to reduce the height of its center of mass, resulting in the blue shape. With inertia, the body will perform a rocking oscillation about the blue shape, damped by internal or frictional dissipation in real situations. With further shape changes, the body may continue rightward with steady or variable velocity \cite{puopolo2016velocity}. Like walking, these rolling motions do essentially no work against friction in the ideal case. They can also be made into stable 3D shapes by adding a mirror-symmetric portion, as for the motion in panel A. The rolling motions can be made to have relatively small changes in curvature, decreasing viscous dissipation.

Many of our computed optima resemble the walking or rolling strategies. The sidewinding motions involve a continuous rolling on and off the ground at certain contact regions, although the body does not assume a circular shape. A small number of other organisms have been found to roll passively (without shape change), including tumbleweed, a type of salamander, a spider, and a woodlouse \cite{armour2006rolling}. Organisms that use active rolling to locomote include a type of shrimp and a caterpillar \cite{armour2006rolling}. There may be various reasons why sidewinding is preferable to a more vertical rolling configuration for snakes. Sidewinding may allow for a better view of the snake's surroundings including prey and predators. In the sidewinding motion, the snake is closer to the ground, which is better for stealth when stalking prey or avoiding predators \cite{lillywhite2014snakes}. Body configurations that are closer to the ground are often more gravitationally stable. Another advantage is that less work is done is lifting portions of the body high off the ground. Although gravitational potential energy can be recovered in a periodic motion, there may be more losses during large conversions of potential to kinetic energy \cite{biewener2018animal}. 

Another possible advantage of sidewinding over the rolling motions in figure \ref{fig:WalkingRollingFig} is that it preserves the snake's upright orientation.  Snakes have an upright orientation in most studies of locomotion over approximately flat surfaces, and their belly scales seem to be adapted for contact with surfaces \cite{lillywhite2014snakes}. One notable and unusual case of an upside-down posture is a specialized antipredator behavior of the hognose snake \cite{hemken1974defensive}.  Upright postures are also preferred by many other organisms such as fish \cite{eidietis2002relative,tasoff2017why}. It is interesting that sidewinding optima occur in our computations even though our model omits biological considerations beyond mechanical efficiency, as well as details of snake physiology that may favor certain body postures such as sidewinding. Sidewinding may be the most efficient rolling type of motion that fits the constraint of small-to-moderate lifting. It is also interesting that sidewinding optima occur only with a very small viscous damping constant. However, the more-efficient alternating lifting or ``walking" optima are also optimal in many cases with large viscous dissipation, which include high-speed motions.

\section{Conclusions \label{sec:conclusions}}

We have developed a model and computational method to find 3D motions that optimize the mechanical efficiency of snake-like locomotion with small-to-moderate lifting off of the ground. The key physical parameters are the two ratios of friction coefficients ($\mu_n/\mu_f, \mu_b/\mu_f$), the lifting amplitude $\tau_{amp}$, and the viscous damping parameter $c_v$. Our stochastic population-based optimization method finds motions that minimize the average input power $\tilde{P}$ while locomoting with very small net rotation per period. 
The same types of optima are found when we vary the numbers of modes in the Chebyshev-Fourier basis that describe the body shapes, so the results are robust with respect to changing the numbers of modes. Although we have mainly focused on the best computed optima, in most cases the second-best computed optimum is very similar to the best, so the algorithm finds the same optima from different initializations. Presumably the best computed optimum has a sizeable basin of attraction and is not very difficult to locate.

For planar (and nonplanar) locomotion, the transition from negligible to dominant viscous damping occurs near $c_v = 10^{-3}$. The optima in the planar case with zero viscous damping are very similar to those computed in \cite{AlbenSnake2013} with a Newton-based optimization method. With nonzero viscous damping, the body shapes are often similar but smoother, and achieve larger displacements per period.

When the allowed lifting amplitude $\tau_{amp}$ is increased from zero, the optimal motions change dramatically. For $\mu_n/\mu_f \leq 3$,
a motion with an s-shaped body and alternating lifting of the midbody and the ends is typical, though there are other types of motions at certain friction coefficients. For $\mu_n/\mu_f  = 10$, the optimal motions are a combination of lifting with curling and sliding at zero $c_v$. At large $c_v$, the motions are a combination of lifting with large amplitude bending and sliding, and somewhat resemble lateral undulation but combined with lifting that alternates between the ends. Other types of optima are seen with moderate and large damping and small or large $\mu_n/\mu_f$ and $\mu_b/\mu_f > 1$, including repeated curling and uncurling with the contacts almost fixed and oriented in the direction of minimal friction.

At all friction coefficient values, optima that resemble sidewinding are seen, but these underperform the optima with an s-shaped body and alternating lifting. The sidewinding optima have a unidirectional wave of contact synchronized with a wave of curvature, tracing out a contiguous contact region (or ``track") on the ground. At different friction coefficient ratios, the body orientation in the lifted region varies. 
Finally, we discussed the resemblance of the computed optima to theoretical walking and rolling motions with essentially zero frictional dissipation. Future work may consider additional physical effects such as a nonplanar substrate, as well as the possible benefits of passive flexibility on efficient nonplanar locomotion \cite{wang2018dynamics}.

\begin{acknowledgments}
This research was supported by the NSF Mathematical Biology program under
award number DMS-1811889.
\end{acknowledgments}


\begin{thebibliography}{10}

\bibitem{shine2003aquatic}
Richard Shine, Harold~G Cogger, Robert~R Reed, Sohan Shetty, and Xavier Bonnet.
\newblock {Aquatic and terrestrial locomotor speeds of amphibious sea-snakes
  (Serpentes, Laticaudidae)}.
\newblock {\em Journal of Zoology}, 259(3):261--268, 2003.

\bibitem{socha2002kinematics}
John~J Socha.
\newblock {Kinematics: Gliding flight in the paradise tree snake}.
\newblock {\em Nature}, 418(6898):603--604, 2002.

\bibitem{hirosebiologically}
S~Hirose.
\newblock {\em {Biologically Inspired Robots: Snake-Like Locomotors and
  Manipulators}}.
\newblock Oxford University Press, 1993.

\bibitem{transeth2009survey}
Aksel~Andreas Transeth, Kristin~Ytterstad Pettersen, and P{\aa}l Liljeb{\"a}ck.
\newblock A survey on snake robot modeling and locomotion.
\newblock {\em Robotica}, 27(07):999--1015, 2009.

\bibitem{hopkins2009survey}
James~K Hopkins, Brent~W Spranklin, and Satyandra~K Gupta.
\newblock {A survey of snake-inspired robot designs}.
\newblock {\em Bioinspiration \& Biomimetics}, 4(2):021001, 2009.

\bibitem{liljeback2012snake}
Pal Liljeb{\"a}ck, Kristin~Ytterstad Pettersen, {\O}yvind Stavdahl, and
  Jan~Tommy Gravdahl.
\newblock {\em Snake Robots: Modelling, Mechatronics, and Control}.
\newblock Springer, 2012.

\bibitem{HaCh2010b}
R~L Hatton and H~Choset.
\newblock {Generating gaits for snake robots: annealed chain fitting and
  keyframe wave extraction}.
\newblock {\em Autonomous Robots}, 28(3):271--281, 2010.

\bibitem{astley2015modulation}
Henry~C Astley, Chaohui Gong, Jin Dai, Matthew Travers, Miguel~M Serrano,
  Patricio~A Vela, Howie Choset, Joseph~R Mendelson, David~L Hu, and Daniel~I
  Goldman.
\newblock Modulation of orthogonal body waves enables high maneuverability in
  sidewinding locomotion.
\newblock {\em Proceedings of the National Academy of Sciences},
  112(19):6200--6205, 2015.

\bibitem{fu2020robotic}
Qiyuan Fu and Chen Li.
\newblock Robotic modelling of snake traversing large, smooth obstacles reveals
  stability benefits of body compliance.
\newblock {\em Royal Society open science}, 7(2):191192, 2020.

\bibitem{astley2020side}
Henry~C Astley, Jennifer~M Rieser, Abdul Kaba, Veronica~M Paez, Ian Tomkinson,
  Joseph~R Mendelson, and Daniel~I Goldman.
\newblock Side-impact collision: mechanics of obstacle negotiation in
  sidewinding snakes.
\newblock {\em Bioinsp. \& Biomim.}, 15(6):065005, 2020.

\bibitem{fu2022snakes}
Qiyuan Fu, Henry Astley, and Chen Li.
\newblock Snakes combine vertical and lateral bending to traverse uneven
  terrain.
\newblock {\em Bioinspiration \& Biomimetics}, 2022.

\bibitem{gray1946mechanism}
J~Gray.
\newblock {The mechanism of locomotion in snakes}.
\newblock {\em J. Exp. Biol.}, 23(2):101--120, 1946.

\bibitem{gans1970snakes}
Carl Gans.
\newblock How snakes move.
\newblock {\em Scientific American}, 222(6):82--99, 1970.

\bibitem{jayne1986kinematics}
Bruce~C Jayne.
\newblock {Kinematics of terrestrial snake locomotion}.
\newblock {\em Copeia}, pages 915--927, 1986.

\bibitem{lillywhite2014snakes}
Harvey~B Lillywhite.
\newblock {\em How Snakes Work: Structure, Function and Behavior of the World's
  Snakes}.
\newblock Oxford University Press, 2014.

\bibitem{gans1984slide}
Carl Gans.
\newblock Slide-pushing: A transitional locomotor method of elongate squamates.
\newblock In {\em Symp Zool Soc London}, volume~52, pages 12--26, 1984.

\bibitem{jayne2020defines}
Bruce~C Jayne.
\newblock What defines different modes of snake locomotion?
\newblock {\em Int. Comp. Biol.}, 60(1):156--170, 2020.

\bibitem{ma2001analysis}
Shugen Ma.
\newblock Analysis of creeping locomotion of a snake-like robot.
\newblock {\em Adv. Robot.}, 15(2):205--224, 2001.

\bibitem{sato2002serpentine}
M~Sato, M~Fukaya, and T~Iwasaki.
\newblock Serpentine locomotion with robotic snakes.
\newblock {\em IEEE Cont. Sys. Mag.}, 22(1):64--81, 2002.

\bibitem{chernousko2005modelling}
Felix~L Chernousko.
\newblock Modelling of snake-like locomotion.
\newblock {\em Appl. Math. Comput.}, 164(2):415--434, 2005.

\bibitem{GuMa2008a}
Z~V Guo and L~Mahadevan.
\newblock {Limbless undulatory propulsion on land}.
\newblock {\em PNAS}, 105(9):3179, 2008.

\bibitem{HuNiScSh2009a}
D~L Hu, J~Nirody, T~Scott, and M~J Shelley.
\newblock {The mechanics of slithering locomotion}.
\newblock {\em Proceedings of the National Academy of Sciences}, 106(25):10081,
  2009.

\bibitem{HuSh2012a}
D~L Hu and M~Shelley.
\newblock {Slithering Locomotion}.
\newblock In {\em Natural Locomotion in Fluids and on Surfaces}, pages
  117--135. Springer, 2012.

\bibitem{aguilar2016review}
Jeffrey Aguilar, Tingnan Zhang, Feifei Qian, Mark Kingsbury, Benjamin McInroe,
  Nicole Mazouchova, Chen Li, Ryan Maladen, Chaohui Gong, Matt Travers, Ross~L
  Hatton, Howie Choset, Paul~B Umbanhowar, and Daniel~I Goldman.
\newblock A review on locomotion robophysics: the study of movement at the
  intersection of robotics, soft matter and dynamical systems.
\newblock {\em Reports on Progress in Physics}, 79(11), 2016.

\bibitem{yona2019wheeled}
Tal Yona and Yizhar Or.
\newblock The wheeled three-link snake model: singularities in nonholonomic
  constraints and stick--slip hybrid dynamics induced by coulomb friction.
\newblock {\em Nonlinear Dynamics}, 95(3):2307--2324, 2019.

\bibitem{rieser2021functional}
Jennifer~M Rieser, Jessica~L Tingle, Daniel~I Goldman, Joseph~R Mendelson,
  et~al.
\newblock Functional consequences of convergently evolved microscopic skin
  features on snake locomotion.
\newblock {\em PNAS}, 118(6), 2021.

\bibitem{alben2019efficient}
Silas Alben.
\newblock Efficient sliding locomotion with isotropic friction.
\newblock {\em Phys. Rev. E}, 99(6):062402, 2019.

\bibitem{AlbenSnake2013}
S~Alben.
\newblock {Optimizing snake locomotion in the plane}.
\newblock {\em Proc. Roy. Soc. A}, 469(2159):1--28, 2013.

\bibitem{wang2014optimizing}
Xiaolin Wang, Matthew~T Osborne, and Silas Alben.
\newblock Optimizing snake locomotion on an inclined plane.
\newblock {\em Phys. Rev. E}, 89(1):012717, 2014.

\bibitem{JiAl2013}
F~Jing and S~Alben.
\newblock {Optimization of two- and three-link snake-like locomotion}.
\newblock {\em Phys. Rev. E}, 87(2):022711, 2013.

\bibitem{alben2020intermittent}
Silas Alben and Connor Puritz.
\newblock Intermittent sliding locomotion of a two-link body.
\newblock {\em Phys. Rev. E}, 101(5):052613, 2020.

\bibitem{alben2021efficient}
Silas Alben.
\newblock Efficient sliding locomotion of three-link bodies.
\newblock {\em Phys. Rev. E}, 103(4):042414, 2021.

\bibitem{childress1981mechanics}
Stephen Childress.
\newblock {\em {Mechanics of swimming and flying}}.
\newblock Cambridge University Press, 1981.

\bibitem{sparenberg1994hydrodynamic}
JA~Sparenberg.
\newblock {\em {Hydrodynamic Propulsion and Its Optimization:(Analytic
  Theory)}}, volume~27.
\newblock Kluwer Academic Pub, 1994.

\bibitem{alben2008optimal}
Silas Alben.
\newblock Optimal flexibility of a flapping appendage in an inviscid fluid.
\newblock {\em J. Fluid Mech.}, 614:355--380, 2008.

\bibitem{alben2009swimming}
Silas Alben.
\newblock On the swimming of a flexible body in a vortex street.
\newblock {\em J. Fluid Mech.}, 635:27--45, 2009.

\bibitem{marvi2014sidewinding}
Hamidreza Marvi, Chaohui Gong, Nick Gravish, Henry Astley, Matthew Travers,
  Ross~L Hatton, Joseph~R Mendelson, Howie Choset, David~L Hu, and Daniel~I
  Goldman.
\newblock Sidewinding with minimal slip: Snake and robot ascent of sandy
  slopes.
\newblock {\em Science}, 346(6206):224--229, 2014.

\bibitem{zhang2021friction}
Xiaotian Zhang, Noel Naughton, Tejaswin Parthasarathy, and Mattia Gazzola.
\newblock Friction modulation in limbless, three-dimensional gaits and
  heterogeneous terrains.
\newblock {\em Nature communications}, 12(1):1--8, 2021.

\bibitem{chong2022moving}
Baxi Chong, Tianyu Wang, Bo~Lin, Shengkai Li, Grigoriy Blekherman, Howie
  Choset, and Daniel~I Goldman.
\newblock Moving sidewinding forward: optimizing contact patterns for limbless
  robots via geometric mechanics.
\newblock In {\em Robotics: science and systems}, 2022.

\bibitem{alexander1996optima}
R~McNeill Alexander.
\newblock {\em Optima for animals}.
\newblock Princeton University Press, 1996.

\bibitem{langerhans2010ecology}
R~Brian Langerhans and David~N Reznick.
\newblock Ecology and evolution of swimming performance in fishes: predicting
  evolution with biomechanics.
\newblock {\em Fish locomotion: an eco-ethological perspective}, pages
  200--248, 2010.

\bibitem{GuggenDG}
HW~Guggenheimer.
\newblock {\em {Differential Geometry}}.
\newblock Dover Publications, 2012.

\bibitem{MaHu2012a}
Hamidreza Marvi and David~L Hu.
\newblock Friction enhancement in concertina locomotion of snakes.
\newblock {\em Journal of The Royal Society Interface}, 9(76):3067--3080, 2012.

\bibitem{schultz2002power}
William~W Schultz and Paul~W Webb.
\newblock Power requirements of swimming: do new methods resolve old questions?
\newblock {\em Integrative and Comparative Biology}, 42(5):1018--1025, 2002.

\bibitem{wang2018dynamics}
Xiaolin Wang and Silas Alben.
\newblock Dynamics and locomotion of flexible foils in a frictional
  environment.
\newblock {\em Proc. R. Soc. A}, 474(2209):20170503, 2018.

\bibitem{o2017modeling}
Oliver~M O'Reilly.
\newblock {\em Modeling nonlinear problems in the mechanics of strings and
  rods}.
\newblock Springer, 2017.

\bibitem{linn2013geometrically}
Joachim Linn, Holger Lang, and Andrey Tuganov.
\newblock Geometrically exact cosserat rods with kelvin--voigt type viscous
  damping.
\newblock {\em Mechanical Sciences}, 4(1):79--96, 2013.

\bibitem{love1892treatise}
AEH Love.
\newblock {\em A Treatise on the Mathematical Theory of Elasticity}.
\newblock Cambridge University Press, 1892.

\bibitem{TaHo2007a}
D~Tam and A~E Hosoi.
\newblock {Optimal stroke patterns for Purcell's three-link swimmer}.
\newblock {\em Phys. Rev. Lett.}, 98(6):68105, 2007.

\bibitem{hatton2013geometric}
Ross~L Hatton, Yang Ding, Howie Choset, and Daniel~I Goldman.
\newblock Geometric visualization of self-propulsion in a complex medium.
\newblock {\em Phys. Rev. letters}, 110(7):078101, 2013.

\bibitem{gutman2015symmetries}
Emiliya Gutman and Yizhar Or.
\newblock {Symmetries and gaits for Purcell's three-link microswimmer model}.
\newblock {\em IEEE Transactions on Robotics}, 32(1):53--69, 2015.

\bibitem{wu2020variation}
Weibin Wu, Shudong Yu, Paul Schreiber, Antje Dollmann, Christian Lutz,
  Guillaume Gomard, Christian Greiner, and Hendrik H{\"o}lscher.
\newblock Variation of the frictional anisotropy on ventral scales of snakes
  caused by nanoscale steps.
\newblock {\em Bioinspiration \& biomimetics}, 15(5):056014, 2020.

\bibitem{schneck1992mechanics}
Daniel~J Schneck.
\newblock {\em Mechanics of muscle}.
\newblock NYU Press, 1992.

\bibitem{cheng1998continuous}
J-Y Cheng, TJ~Pedley, and JD~Altringham.
\newblock A continuous dynamic beam model for swimming fish.
\newblock {\em Phil. Trans. Roy. Soc. London B}, 353(1371):981--997, 1998.

\bibitem{al2018biomechanics}
Adil Al~Mayah.
\newblock {\em Biomechanics of soft tissues: principles and applications}.
\newblock CRC Press, 2018.

\bibitem{HaBuHoCh2011a}
R~L Hatton, L~J Burton, A~E Hosoi, and H~Choset.
\newblock {Geometric maneuverability with applications to low Reynolds number
  swimming}.
\newblock In {\em Intelligent Robots and Systems (IROS), 2011 IEEE/RSJ
  International Conference on}, pages 3893--3898. IEEE, 2011.

\bibitem{sastra2009dynamic}
Jimmy Sastra, Sachin Chitta, and Mark Yim.
\newblock Dynamic rolling for a modular loop robot.
\newblock {\em The International Journal of Robotics Research}, 28(6):758--773,
  2009.

\bibitem{armour2006rolling}
Rhodri~H Armour and Julian~FV Vincent.
\newblock Rolling in nature and robotics: a review.
\newblock {\em Journal of Bionic Engineering}, 3(4):195--208, 2006.

\bibitem{puopolo2016velocity}
Michael~G Puopolo and Jamey~D Jacob.
\newblock Velocity control of a cylindrical rolling robot by shape changing.
\newblock {\em Advanced Robotics}, 30(23):1484--1494, 2016.

\bibitem{li2021electrically}
Wen-Bo Li, Wen-Ming Zhang, Qiu-Hua Gao, Qiwei Guo, Song Wu, Hong-Xiang Zou,
  Zhi-Ke Peng, and Guang Meng.
\newblock Electrically activated soft robots: Speed up by rolling.
\newblock {\em Soft Robotics}, 8(5):611--624, 2021.

\bibitem{shah2021soft}
Dylan~S Shah, Joshua~P Powers, Liana~G Tilton, Sam Kriegman, Josh Bongard, and
  Rebecca Kramer-Bottiglio.
\newblock A soft robot that adapts to environments through shape change.
\newblock {\em Nature Machine Intelligence}, 3(1):51--59, 2021.

\bibitem{biewener2018animal}
Andrew Biewener and Sheila Patek.
\newblock {\em Animal locomotion}.
\newblock Oxford University Press, 2018.

\bibitem{hemken1974defensive}
Brenda~S Hemken.
\newblock {\em Defensive Behavior of the Hognose Snake (Heterodon
  platyrhinos)}.
\newblock PhD thesis, Eastern Illinois University, 1974.

\bibitem{eidietis2002relative}
L~Eidietis, TL~Forrester, and PW~Webb.
\newblock Relative abilities to correct rolling disturbances of three
  morphologically different fish.
\newblock {\em Canadian journal of zoology}, 80(12):2156--2163, 2002.

\bibitem{tasoff2017why}
Harrison Tasoff.
\newblock Why don't fish swim upside down?
\newblock {\em Hakai Magazine}, November 2017.

\end{thebibliography}
\end{document}